\documentclass[
aps,
prd,
preprint,
nofootinbib,
preprintnumbers,
superscriptaddress,
longbibliography
]{revtex4-1}
\usepackage{graphicx}
\usepackage{float}
\usepackage{caption}
\usepackage{subcaption}
\usepackage{dcolumn}
\usepackage{bm}
\usepackage{amsmath}
\usepackage{amsfonts} 
\usepackage{latexsym}
\usepackage{bbm}
\usepackage{color}
\usepackage{amssymb}
\usepackage{amsthm}
\usepackage{appendix}
\usepackage{mathtools}
\usepackage{cases}
\usepackage{multirow}

\usepackage[thicklines]{cancel}
\usepackage[
    colorlinks=true,
    citecolor=blue,
    linkcolor=blue,
    urlcolor=blue
]{hyperref}
\begin{document}
	
\title{Precise bubble wall velocity in a specific phase transition pattern in the CxSM and beyond }
	
\author{Shihang Tang}

\author{Fa Peng Huang}
\email{Corresponding Author. 
 huangfp8@sysu.edu.cn}

\author{ Chikako Idegawa}%

\affiliation{MOE Key Laboratory of TianQin Mission, TianQin Research Center for
Gravitational Physics \& School of Physics and Astronomy, Frontiers
Science Center for TianQin, Gravitational Wave Research Center of CNSA, 
Sun Yat-sen University (Zhuhai Campus), Zhuhai 519082, China}

\bigskip
	
\date{\today}
	
\begin{abstract}
The bubble wall velocity is a key quantity in cosmological first-order phase transitions, with important implications for electroweak baryogenesis, gravitational wave signals, the dark matter relic density and primordial black holes formed during the phase transition. However, it is often treated as a free input in phenomenological studies, while a self-consistent determination remains challenging. In this work, taking the complex singlet extension of the Standard Model as an example, we investigate the bubble wall dynamics and velocity in a specific electroweak phase transition pattern where both the Higgs field and the coupled singlet scalar experience friction. The microscopic friction arising from particle interactions with the plasma is evaluated using Boltzmann transport equations, while the macroscopic plasma response is described through hydrodynamic analysis. By applying the steady state force balance condition, we numerically determine the bubble wall velocity for different model parameters. We show that the wall velocity is governed by the competition between the driving force from the effective potential and plasma friction, and that its variation can significantly affect the baryon asymmetry. 
Our study provides a quantitative investigation of bubble wall dynamics in this overlooked phase transition pattern and its implications for early Universe phenomenology.
\end{abstract}

\maketitle

\section{Introduction}\label{sec:intro}

The electroweak phase transition (EWPT) provides a theoretically well motivated connection between particle physics and the thermal history of the early Universe. Electroweak baryogenesis (EWBG) is a testable mechanism for generating the observed baryon asymmetry. It requires baryon number violation, CP-violating interactions, and a departure from thermal equilibrium. A sufficiently strong first-order electroweak phase transition (FOEWPT) is one of the key ingredients required for EWBG~\cite{Kuzmin:1985mm,Funakubo:1996dw,cline_baryogenesis_2021,grinstein_electroweak_2008,Morrissey:2012db,vandeVis:2025efm}. The requirement of a sufficiently strong FOEWPT is not satisfied in the Standard Model (SM), whose EWPT is a crossover according to nonperturbative studies~\cite{Kajantie:1996mn,Csikor:1998eu}. This motivates extensions of the scalar sector in which the thermal effective potential can support a FOEWPT. The complex singlet extension of the Standard Model (CxSM) offers a simple and well studied framework in which a FOEWPT can be realized, and several viable CxSM scenarios have been investigated~\cite{Barger:2008jx,cho_electroweak_2021,Cho:2022zfg,Ghosh:2025rbt}. A FOEWPT proceeds through the nucleation and expansion of bubbles of the broken phase in the surrounding plasma, making bubble wall dynamics an essential part of the transition~\cite{Rubakov:1996vz,kozaczuk_bubble_2015,Hindmarsh:2020hop}.

The bubble wall velocity $v_w$ is a key dynamical parameter of a FOEWPT. Together with the wall thickness, it sets the time and length scales over which particles experience the varying scalar background as they cross the wall, directly affecting the transport processes relevant to EWBG~\cite{dorsch_wall_2021,Cline:2020jre,cline_baryogenesis_2021}. A FOEWPT can also generate stochastic gravitational wave (GW) backgrounds through bubble collisions, sound waves, plasma turbulence, and other proposed mechanisms~\cite{Kosowsky:1991ua,Kosowsky:1992vn,Hindmarsh:2013xza,Kamionkowski:1993fg,Chang:2025rda,Qiu:2025tmn}; next generation space based detectors such as TianQin, Taiji, and LISA could probe these signals~\cite{Luo:2025ewp,TianQin:2015yph,Hu:2017mde,LISA:2017pwj,Yang:2025ybx}. The wall velocity significantly affects the peak frequency and amplitude of the GW spectrum~\cite{Huang:2023eal,Yang:2024npd,Yang:2025ybx,Caprini:2019egz}; it also influences heavy dark matter production and its relic abundance~\cite{Baker:2019ndr,Chway:2019kft,Huang:2017kzu,Azatov:2021ifm,Azatov:2022tii}, and can affect primordial black hole formation in scenarios where particles reflected from the advancing walls accumulate into overdensities~\cite{Baker:2021nyl,Baker:2021sno}. Thus, a self-consistent determination of $v_w$, rather than treating it as an independent phenomenological input, is important for obtaining a consistent description of the associated early Universe phenomenology~\cite{Athron:2023xlk,Caprini:2019egz,Baker:2019ndr,Chway:2019kft,Baker:2021nyl}.

In the steady-state regime, the wall velocity follows from the scalar field equations of motion (EOMs): the thermodynamic pressure that drives the expansion must be balanced against the friction generated by the plasma, where the plasma response enters through both nonequilibrium transport and hydrodynamic backreaction~\cite{Arnold:1993wc,Moore:1995ua,Moore:1995si,Megevand:2013hwa}. Early studies used simplified treatments of wall friction and plasma dynamics~\cite{Dine:1992wr,Liu:1992tn,Ignatius:1993qn}. Seminal calculations by Moore and Prokopec showed how out of equilibrium particle distributions generate friction, and treated these departures by solving the Boltzmann equations within a specified fluid ansatz (the fluid approximation)~\cite{Moore:1995ua,Moore:1995si}; Moore later obtained analytic results at leading logarithmic order~\cite{Moore:2000wx}. Bödeker and Moore established a criterion for deciding whether plasma friction can bring a bubble wall to terminal velocity or whether it undergoes runaway acceleration~\cite{Bodeker:2009qy}. Later work developed more systematic treatments that combine scalar field dynamics with particle transport and the plasma response~\cite{Konstandin:2014zta,Laurent:2022jrs,DeCurtis:2022hlx}.

Microscopic transport methods compute the friction due to departures of the particle distribution functions from equilibrium. In the semiclassical approach, these departures are obtained by solving Boltzmann equations for the relevant particle species, with collision terms describing the interactions that drive the distributions toward equilibrium. Progress in this direction includes treatments beyond the leading-log approximation and numerical methods for evaluating the collision integrals~\cite{Moore:1995ua,Moore:1995si,Moore:2000wx,John:2000zq,Wang:2020zlf,Konstandin:2014zta,DeCurtis:2022hlx,DeCurtis:2023hil,DeCurtis:2024hvh}. 
These integrals remain computationally demanding to evaluate and may be sensitive to soft, infrared-enhanced scattering processes and to the treatment of thermal screening. Spectral methods have enabled more systematic treatments of the coupled scalar field, transport, and fluid equations~\cite{DeCurtis:2023hil,ekstedt_how_2024,Branchina:2025adj}. Recent studies have also quantified uncertainties associated with the particle species included in the nonequilibrium treatment, thermal masses, collision integrals, wall profile ansatze, and the effective potential~\cite{Hirvonen:2021zej,Lofgren:2021ogg,vandeVis:2025plm,Branchina:2026nql}. Complementary field theoretic analyses have examined the force on the wall beyond conventional kinetic treatments and identified additional dissipative contributions from particle production in certain regimes~\cite{BarrosoMancha:2020fay,Ai:2025bjw}.

The hydrodynamic description accounts for the macroscopic response of the plasma to the moving bubble wall. Energy momentum conservation and matching conditions across the phase boundary determine the possible fluid profiles and lead to deflagration, detonation, and hybrid solutions~\cite{steinhardt_relativistic_1982,Ignatius:1993qn,Megevand:2009gh,espinosa_energy_2010,Dorsch:2023tss}.
This response can modify the temperature and fluid velocity profiles around the wall, thereby impeding its propagation. In particular, hydrodynamic obstruction may prevent further acceleration even when microscopic friction alone would allow it
~\cite{konstandin_hydrodynamic_2011,wang_hydrodynamic_2023,Branchina:2025jou}. The existence of a steady-state solution, however, does not guarantee that an expanding bubble will reach it. Real-time simulations under the local thermal equilibrium (LTE) assumption have shown that early-time acceleration can drive walls to supersonic velocities before a heated fluid shell forms. This behavior is not captured by approaches that assume stationary solutions throughout the evolution~\cite{Krajewski:2024gma}. Studies based on the LTE and ballistic limits have also given bounds on the wall velocity without requiring a detailed evaluation of the collision terms~\cite{Ai:2024btx}. A steady-state wall velocity calculation must therefore be interpreted within the dynamical regime and under the plasma assumptions for which it was constructed.

These issues become more involved when several scalar backgrounds vary across the wall. In practical wall velocity calculations, the scalar profiles may be simplified by focusing on a dominant order parameter or by adopting a prescribed ansatz for the trajectory in field space; such choices can limit the treatment of transitions in which several backgrounds change appreciably~\cite{Dorsch:2018pat,vandeVis:2025plm}. For particle species whose masses depend on more than one background field, the nonequilibrium force contributes to the corresponding scalar field equations through derivatives of the field dependent masses. The CxSM provides a concrete setting in which to examine this effect: FOEWPT have been studied in the nearly degenerate scalar scenario~\cite{cho_electroweak_2021,Cho:2022zfg}. In the particular phase transition pattern considered here, both the Higgs and singlet backgrounds vary across the wall, while the singlet background remains nonzero in both phases. Its variation can therefore contribute to the friction terms in the coupled scalar field EOMs. The phase transition pattern studied here thus requires the friction to be evaluated along the coupled Higgs singlet profile, rather than along the Higgs direction alone.

In this work, we determine the steady-state wall velocity for this CxSM transition pattern by combining microscopic particle transport with the hydrodynamic response of the plasma. We calculate the collision terms for the selected particle species and solve the corresponding Boltzmann equations together with the hydrodynamic equations for the plasma. We then impose two steady-state moment conditions on the coupled scalar field EOMs to determine the wall velocity and wall thickness. We apply this procedure to benchmark points in the nearly degenerate scalar scenario~\cite{Abe:2021nih,cho_electroweak_2021,Cho:2022zfg} and examine the dependence of the calculated velocities on the phase transition strength. Over the parameter range considered, the numerical results are well described by an empirical fit as a function of the phase transition strength $\alpha_n$. This fit provides a convenient parametrization of the scan, but it is not a universal relation: the wall velocity depends on the model dependent driving force, microscopic friction, and hydrodynamic response. We also estimate the baryon asymmetry using the self-consistently calculated wall velocities and the associated nonequilibrium chemical potential perturbations. For comparison, we consider a conventional estimate in which the wall velocity is varied as a free input while the other quantities entering the estimate are held fixed at their values for a reference benchmark point. The comparison illustrates the importance of determining the wall velocity self-consistently, since varying it independently does not preserve the correlations among the wall velocity, the hydrodynamic plasma state, and the particle perturbations. Our results provide a quantitative study of this phase transition pattern and illustrate how the framework could be extended to other models with coupled scalar backgrounds.

The paper is organized as follows. In Sec.~\ref{sec:CxSM}, we introduce the CxSM,  present the finite temperature effective potential, and discuss the relevant thermodynamic properties of the FOEWPT. Sec.~\ref{sec:bubble_wall_dynamics} constructs the hydrodynamic and microscopic transport framework used to determine the steady-state wall velocity. The numerical results for the benchmark parameter space are given in Sec.~\ref{sec:numerical_results}, while Sec.~\ref{sec:baryon} discusses the impact of the calculated wall velocity on the baryon asymmetry. Conclusions and discussions are given in Sec.~\ref{sec:conclusion}. Additional details of the scattering matrix elements,  the phase space integration, and the numerical collision coefficients are collected in the appendices.

\section{The Complex Singlet Extension of the Standard Model}
\label{sec:CxSM}

To study the FOEWPT and associated bubble wall dynamics beyond the SM, we consider the CxSM. The model extends the scalar sector of the SM by introducing a complex gauge-singlet scalar field. 
In addition to providing a simple framework for studying physics beyond the SM, the CxSM can accommodate cosmological phenomena such as the dark matter abundance and the baryon asymmetry of the Universe, which cannot be explained within the SM. In this section, we introduce the scalar sector of the CxSM and construct the finite temperature effective potential used in the subsequent analysis of the FOEWPT and bubble wall dynamics.

\subsection{The finite temperature effective potential}
\label{subsec:effective_potential}

The Lagrangian density of the CxSM can be written as~\cite{Barger:2008jx} 
\begin{equation}
  \mathcal{L} = \mathcal{L}_g+\mathcal{L}_f+\mathcal{L}_H+\mathcal{L}_Y\,,
\end{equation}
where $\mathcal{L}_g$ and $\mathcal{L}_f$ denote the gauge and fermion sectors, respectively, while the scalar and Yukawa sectors are represented by $\mathcal{L}_H$ and $\mathcal{L}_Y$. The scalar sector differs from that of the SM by the introduction of a complex gauge-singlet field $S$,
\begin{equation}
  \mathcal{L}_H = -(D_\mu H)^\dagger (D^\mu H) - \partial_\mu S^*\partial^\mu S - V_0(H,\,S) \,,
\end{equation}
where $H$ denotes the SM Higgs doublet and $V_0(H,\,S)$ is the tree-level scalar potential. To investigate the vacuum structure and the FOEWPT, quantum and thermal corrections to the tree-level potential must be taken into account. We therefore construct the finite temperature effective potential by including the one-loop Coleman-Weinberg correction, the corresponding counterterm contribution, and the finite temperature correction.

For definiteness, we consider the CP-conserving CxSM, in which all parameters in the scalar potential are taken to be real. The most general tree-level scalar potential under this assumption can be written as~\cite{Idegawa:2023bkh}
\begin{equation}
  \begin{split}
    V_0(H,S)=&\frac{m^2}{2}H^\dagger H+\frac{\lambda}{4}(H^\dagger H)^2+\frac{b_2}{2}|S|^2+\frac{d_2}{4}|S|^4+\frac{\delta_2}{2}H^\dagger H|S|^2\\
    &+\left(a_1S+\frac{b_1}{4}S^2+h.c.\right) \,.
    \label{eq:primordial_tree_potential}
  \end{split}
\end{equation}
In general, $a_1$ and $b_1$ can be complex. A phase redefinition of the singlet field can be used to remove one of their phases. In the CP-conserving scenario considered here, we take $a_1$ and $b_1$ to be real. We parametrize the scalar fields in terms of their background components as
\begin{equation}
  H =\left(0,~\frac{1}{\sqrt{2}}\varphi\right)^T,~~~ S =\frac{1}{\sqrt{2}}\left(\varphi_s+i\varphi_\chi\right)\,,
\end{equation}
where $\varphi$ denotes the neutral Higgs background field, while $\varphi_s$ and $\varphi_\chi$ correspond to the real and imaginary components of the singlet field, respectively. Substituting these expressions into the tree-level potential gives
\begin{equation}
  \begin{split}
    V_0(\varphi, \varphi_s, \varphi_\chi)=&\frac{m^2}{4}\varphi^2+\frac{\lambda}{16}\varphi^4+\sqrt{2}a_1 \varphi_s+\frac{b_2+b_1}{4}\varphi_s^2 \\
    &+\frac{b_2-b_1}{4}\varphi_\chi^2+\frac{d_2}{16}(\varphi_s^2+\varphi_\chi^2)^2+\frac{\delta_2}{8}\varphi^2(\varphi_s^2+\varphi_\chi^2)\,.
    \label{eq:primordial_tree_potential_expanded}
  \end{split}
\end{equation}

The vacuum configurations are obtained by requiring the first derivatives of the potential with respect to the background fields to vanish. In the CP-conserving limit considered here, $\varphi_\chi=0$ is an extremum of the potential. This choice preserves the $\varphi_\chi\rightarrow-\varphi_\chi$ reflection symmetry and avoids the formation of domain walls associated with spontaneous breaking of the corresponding discrete symmetry. We therefore restrict the subsequent analysis to the $\varphi_\chi=0$ subspace. The tree-level potential then reduces to
\begin{equation}
	V_0(\varphi, \varphi_s)=\frac{m^2}{4}\varphi^2+\frac{\lambda}{16}\varphi^4+\sqrt{2}a_1 \varphi_s+\frac{b_2+b_1}{4}\varphi_s^2+\frac{d_2}{16}\varphi_s^4+\frac{\delta_2}{8}\varphi^2\varphi_s^2 \,.
\end{equation}

The field dependent scalar mass matrix is obtained from the second derivatives of the potential with respect to $\varphi$ and $\varphi_s$. Since the two scalar fields mix, the resulting mass matrix is not diagonal in the $(\varphi,\varphi_s)$ basis. The scalar mass matrix is diagonalized by an orthogonal rotation, yielding the field dependent mass eigenvalues
\begin{equation}
  O(\theta)^{-1}\mathcal{M}^2_h O(\theta)=\begin{pmatrix}
    \bar{m}^2_{h1} & 0 \\
    0 & \bar{m}^2_{h2}
  \end{pmatrix}
  ,~~~~
  O(\theta)=\begin{pmatrix}
    \cos\theta & -\sin\theta \\
    \sin\theta & \cos\theta
  \end{pmatrix}
  \,,
\end{equation}
where $\theta$ denotes the scalar mixing angle. The resulting field dependent mass eigenvalues of the two mixed scalar states and the singlet pseudoscalar-like state $\chi$ are
\begin{equation}
  \begin{split}
    \bar{m}^2_{h_{1,\,2}}(\varphi, \varphi_s)&=\frac{1}{2}\left[m_{h}^2+m_{hs}^2\mp\sqrt{\left(m_{h}^2-m_{hs}^2\right)^2+\delta^2_2\varphi^2\varphi_s^2}\right] \,,\\
    \bar{m}^2_{\chi}(\varphi, \varphi_s)&=\frac{b_2-b_1}{2}+\frac{d_2}{4}\varphi_s^2+\frac{\delta_2}{4}\varphi^2 \,,
  \end{split}
\end{equation}
with
\begin{equation}
  m_{h}^2=\frac{m^2}{2}+\frac{3\lambda}{4}\varphi^2+\frac{\delta_2}{4}\varphi_s^2,  ~~  m_{hs}^2=\frac{b_2+b_1}{2}+\frac{\delta_2}{4}\varphi^2+\frac{3d_2}{4}\varphi_s^2  \,.
  \label{eq:Higgs_mass_squre}
\end{equation}

The field dependent masses of selected SM particles are given by
\begin{equation}
  \bar{m}^2_W=\frac{g_2^2}{4}\varphi^2 \,,~ ~~~  \bar{m}^2_Z=\frac{g_2^2+g_1^2}{4}\varphi^2 \,,~ ~~~  \bar{m}^2_t=\frac{y_t^2}{2}\varphi^2 \,,~ ~~~  \bar{m}^2_b=\frac{y_b^2}{2}\varphi^2 \,.
  \label{eq:SM_partical_mass_squre}
\end{equation}
Here, $g_2$ and $g_1$ are the gauge couplings associated with $SU(2)_L$ and $U(1)_Y$, respectively, while $y_t$ and $y_b$ denote the Yukawa couplings of the top and bottom quarks. In the numerical analysis, only the particle species included in the effective potential are retained in the corresponding loop contributions.

The tree-level potential alone is insufficient to describe the temperature dependent vacuum structure. Quantum corrections modify the zero temperature effective potential, while thermal effects alter the relative depths and locations of the minima at finite temperature. We therefore include both contributions in the effective potential.

At one-loop order and zero temperature, the quantum correction is given by the Coleman-Weinberg potential~\cite{Coleman:1973jx},
\begin{equation}
  V_{CW}(\varphi, \varphi_s)=\sum_i\frac{N_i\bar{m}_i^4}{64\pi^2}\left(\ln\frac{\bar{m}_i^2}{\bar{\mu}^2}-c_i\right) \,,
  \label{eq:V_CW}
\end{equation}
where $N_i$ denotes the number of degrees of freedom of the particle species $i$, including the fermionic minus sign. For the species retained in our calculation, we have $N_{h_{1,2},\chi}=1$, $N_W=6$, $N_Z=3$, and $N_t=-12$. The constant $c_i$ depends on the particle species and the renormalization prescription, with $c_i=3/2$ for scalars and fermions and $c_i=5/6$ for gauge bosons in the $\overline{\rm MS}$ scheme. The quantity $\bar{\mu}$ denotes the renormalization scale, which is chosen at the electroweak scale in the numerical analysis.

In the on-shell-like renormalization scheme adopted here, the finite counterterms are chosen such that the one-loop corrections do not shift the tree-level vacuum expectation values and scalar masses~\cite{Anderson:1991zb,Chiang:2018gsn}. The counterterms can then be written as
\begin{equation}
  V_{CT}(\varphi, \varphi_s)=-\sum_{i}\frac{N_i}{64\pi^2}\left[\bar{m}_i^4\left(\ln\frac{m_i^2(\text{vac})}{\bar{\mu}^2}+\frac{3}{2}-c_i\right)-2\bar{m}_i^2m_i^2({\text{vac}})\right] \,.
\end{equation}
Here, $m_i^2(\mathrm{vac})$ denotes the squared mass evaluated at the zero temperature vacuum.

At finite temperature, the thermal contribution to the effective potential is given by
\begin{equation}
	V_T(\varphi, \varphi_s, T)=\sum_i\frac{N_iT^4}{2\pi^2}I_{B,F}\left(\frac{M_i^2}{T^2}\right) \,,
\end{equation}
where the thermal functions are defined as
\begin{equation}
	I_{B,F}(a^2)=\int_0^{\infty}\,dx\, x^2\ln(1\mp e^{-\sqrt{x^2+a^2}}),~~~M_i^2=\bar{m}^2_i(\varphi,\varphi_s)+\Pi_i(T) \,.
\end{equation}
Here, the upper (lower) sign corresponds to bosons (fermions), and $\Pi_i(T)$ denotes the thermal self-energy. The thermal masses are incorporated in the resummed propagators in order to account for the dominant infrared thermal corrections.

Combining the tree-level potential, the one-loop zero temperature corrections, and the finite temperature contribution, the effective potential used throughout this work is given by
\begin{equation}
	V_{\text{eff}}(\varphi, \varphi_s,T)=V_0(\varphi, \varphi_s)+V_{CW}(\varphi, \varphi_s)+V_{CT}(\varphi, \varphi_s)+V_T(\varphi, \varphi_s,T) \,.
\end{equation}
This effective potential determines the temperature dependent vacuum structure and provides the background for the analysis of the FOEWPT presented in the following section.

\subsection{Thermodynamics of the electroweak phase transition}
\label{subsec:PT_dynamics}

The temperature evolution of the finite temperature effective potential determines the vacuum structure and, consequently, the nature and dynamics of the FOEWPT in the early Universe. Based on the effective potential constructed in Sec.~\ref{subsec:effective_potential}, we investigate the phase transition history and introduce the thermodynamic quantities that will be used in the subsequent calculation of the bubble wall velocity.

The benchmark parameter set used in this section is listed in Table~\ref{tab:effective_potential_benchmark}. For this parameter point, the temperature dependent effective potential exhibits a characteristic FOEWPT.

\begin{table}[htbp]
\centering
\caption{Benchmark parameters of the effective potential}
\label{tab:effective_potential_benchmark}
\begin{tabular}{lcccc}
\hline\hline
$\theta~[\text{rad}]$ & $a_1$~[GeV$^3$] & $m^2$~[GeV$^2$] & $b_1$~[GeV$^2$] & $b_2$~[GeV$^2$] \\
\hline
$\pi/3.992$ & $-6576.1720$ & $-(124.5006)^2$ & $(107.6751)^2$ & $-(178.0019)^2$ \\
\hline\hline
$\lambda$ & $d_2$ &$\delta_2$ & $\bar{\mu}$~[GeV]\\
\hline
$0.5114$ & $3.9214$ & $1.6855$ & $246.22$ \\
\hline
\end{tabular}
\end{table}

\begin{figure}[htbp]
  \centering
  \captionsetup[subfigure]{aboveskip=-6pt}  

  \begin{subfigure}{0.45\linewidth}
    \includegraphics[width=\linewidth]{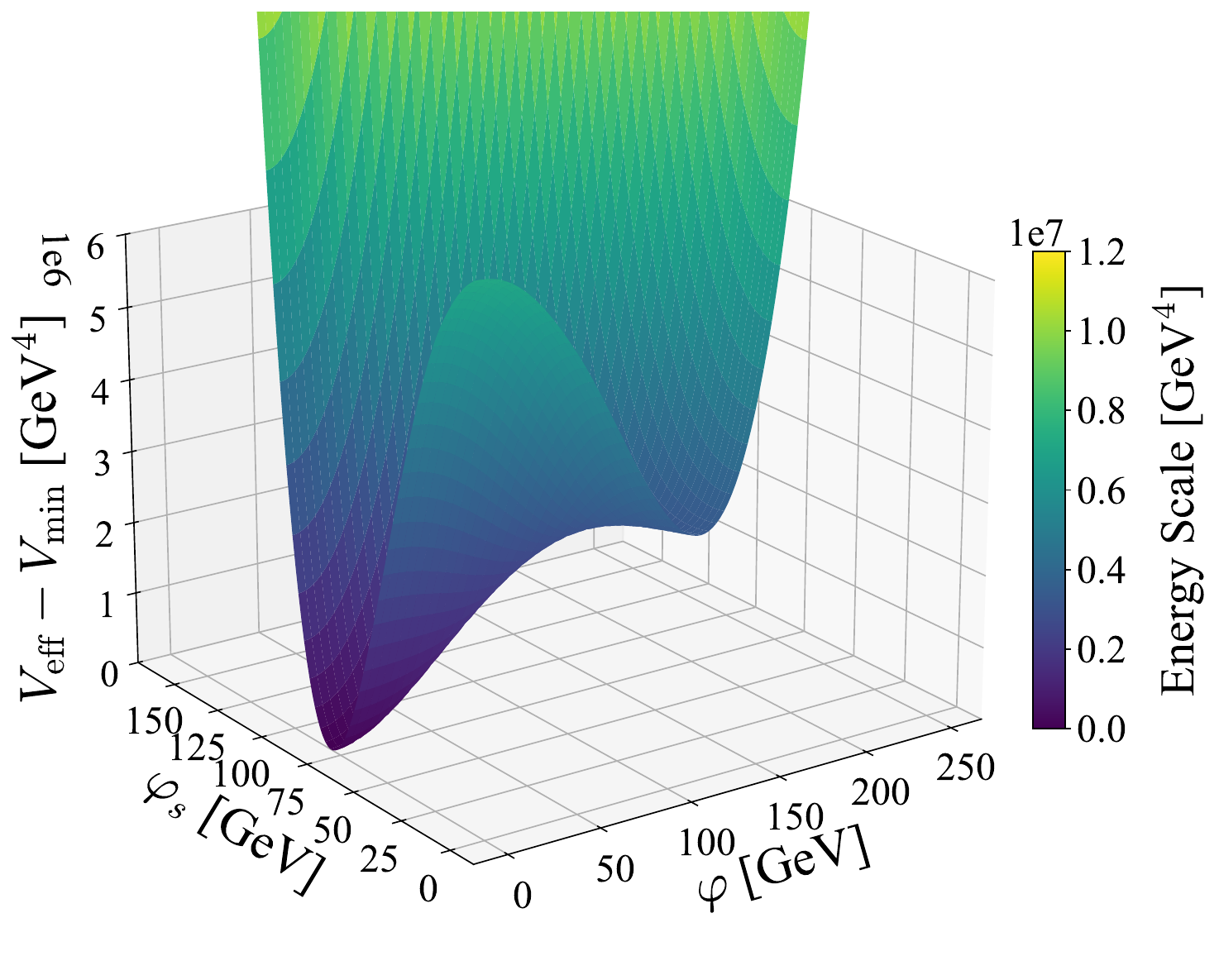}
    \caption{$T>T_c$}
    \label{subfig:T_bigger_Tc}
  \end{subfigure}
  \hfil
  \begin{subfigure}{0.45\linewidth}
    \includegraphics[width=\linewidth]{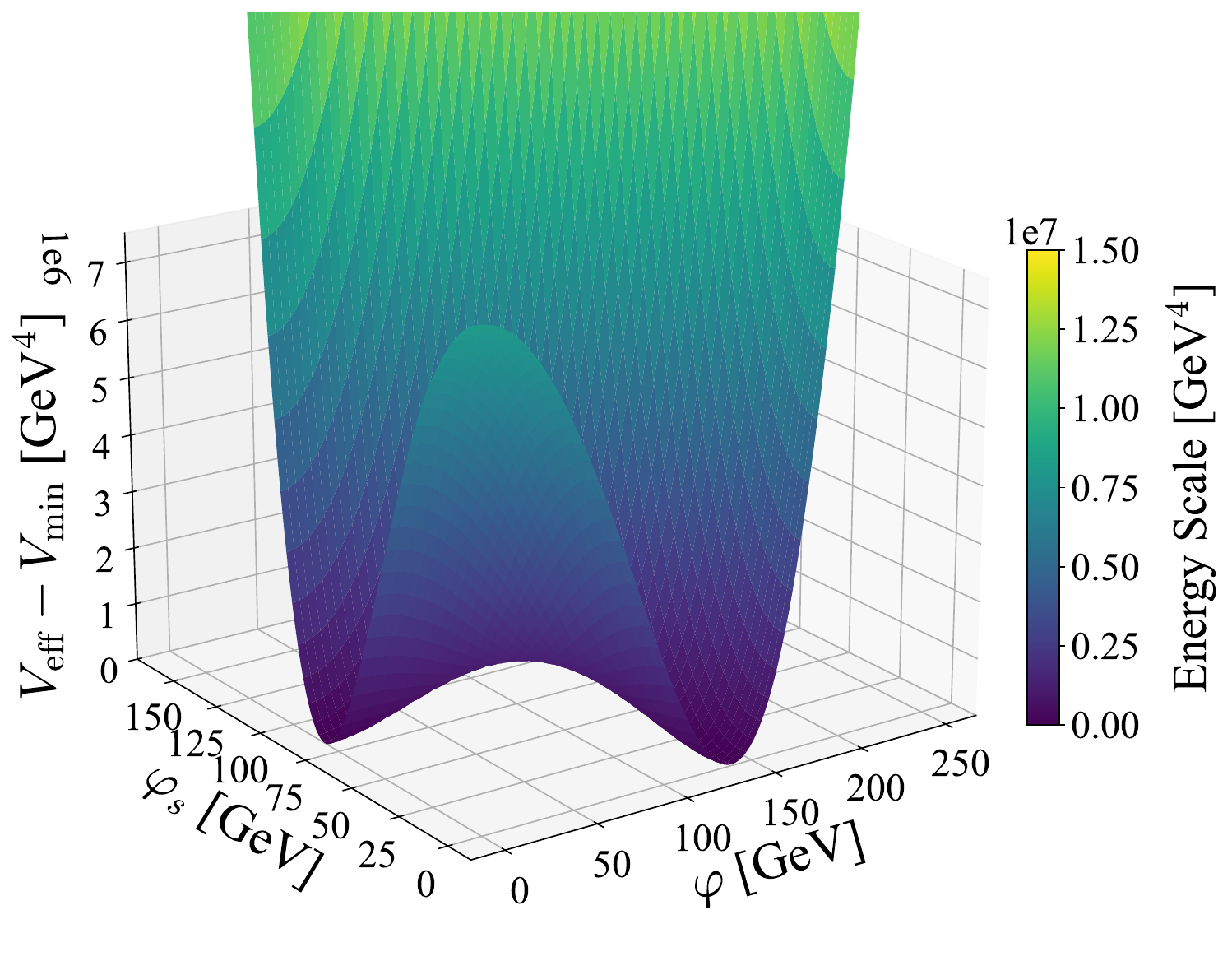}
    \caption{$T=T_c$}
    \label{subfig:T_Tc}
  \end{subfigure}
  
  \begin{subfigure}{0.45\linewidth}
    \includegraphics[width=\linewidth]{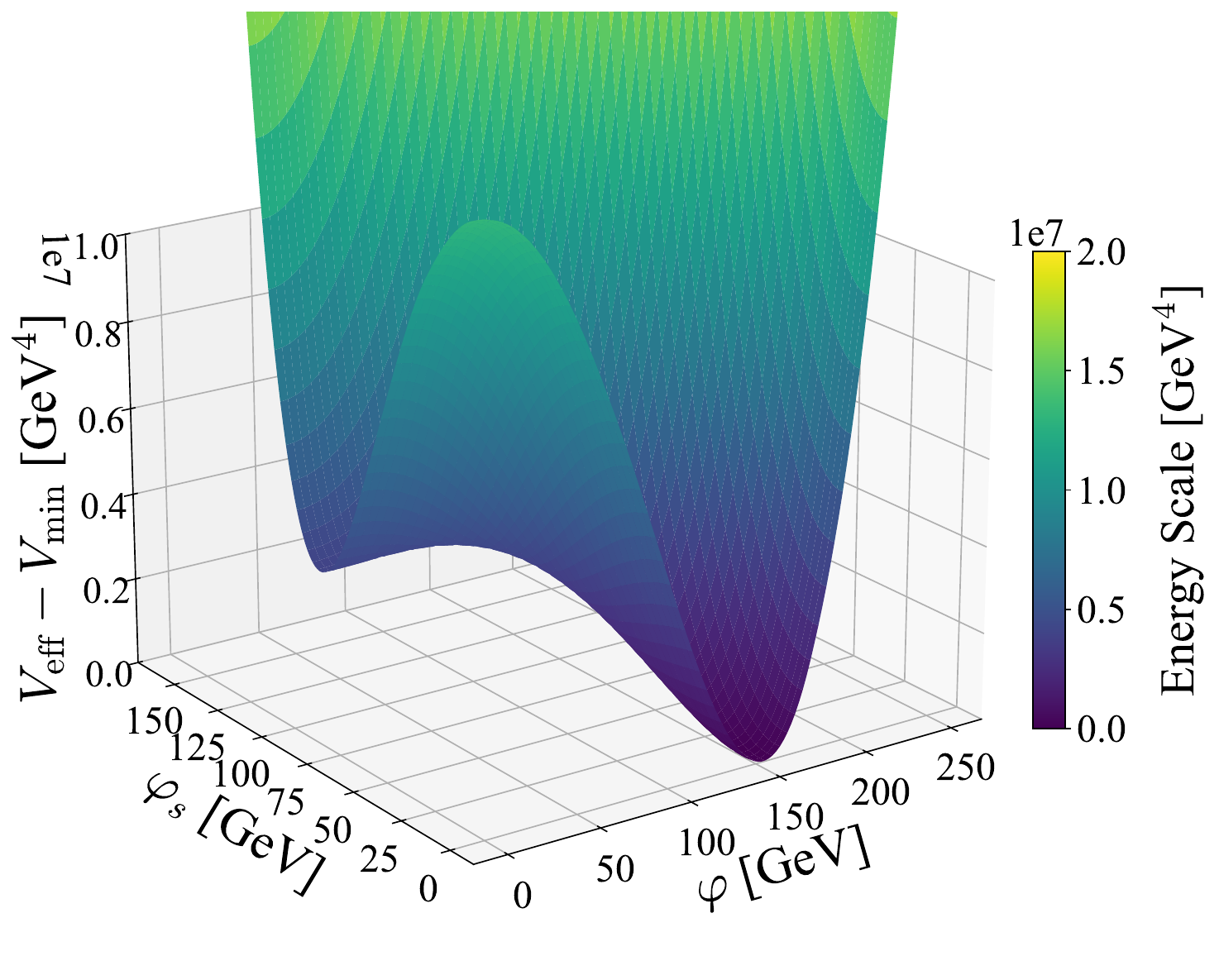}
    \caption{$T<T_c$}
    \label{subfig:T_less_Tc}
  \end{subfigure}
  \hfil
  \begin{subfigure}{0.45\linewidth}
    \includegraphics[width=\linewidth]{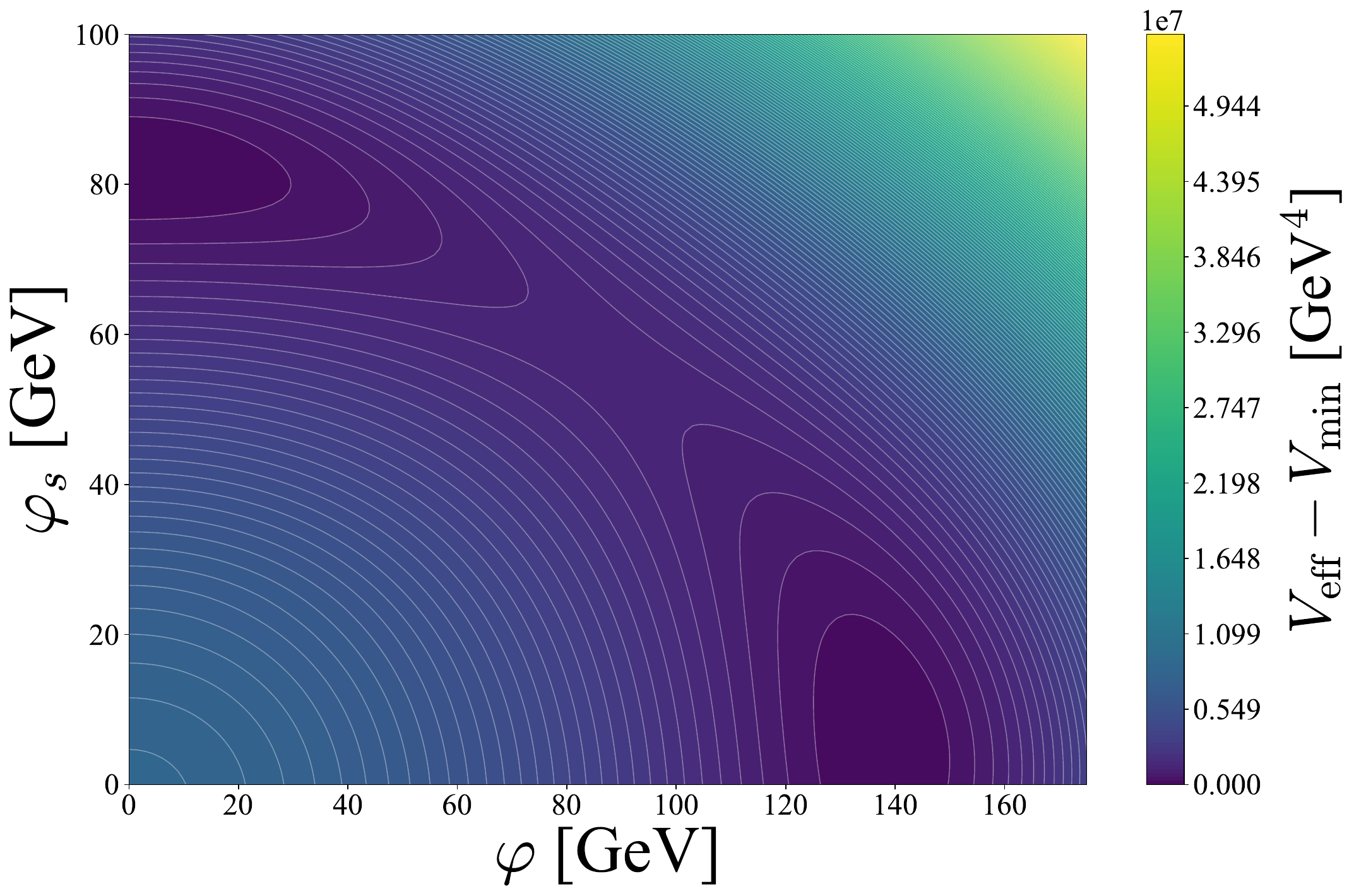}
    \caption{$T=T_c$}
    \label{subfig:T_Tc_2D}
  \end{subfigure}
  
  \caption{Temperature evolution of the effective potential for the benchmark parameter point. For presentation purposes, the effective potential is shifted by an additive constant at each temperature such that its minimum is set to zero. Panel (a) shows the three-dimensional profile of $V_{\mathrm{eff}}(\varphi,\varphi_s,T)$ for $T>T_c$, panel (b) shows the profile at the critical temperature $T=T_c$, and panel (c) shows the profile for $T<T_c$. Panel (d) presents the corresponding two-dimensional contour plot at $T=T_c$ as a complementary view of the three-dimensional potential.}
  \label{fig:CxSM_CPC_JK_Veff}
\end{figure}

Figure~\ref{fig:CxSM_CPC_JK_Veff} illustrates the evolution of $V_{\mathrm{eff}}(\varphi,\varphi_s,T)$ in the $(\varphi,\varphi_s)$ field space for the benchmark parameter point. As shown in Fig.~\ref{subfig:T_bigger_Tc}, for $T>T_c$, the global minimum lies along the $\varphi_s$ direction, while a second local minimum develops near the $\varphi$ axis. At temperatures well above $T_c$, this second minimum is substantially higher in free energy and is therefore thermodynamically disfavored. As the Universe expands and cools, the two minima approach each other in free energy. At the critical temperature $T_c$, they become degenerate, as shown in Fig.~\ref{subfig:T_Tc} and Fig.~\ref{subfig:T_Tc_2D}. Upon further cooling, the minimum near the $\varphi$ axis becomes energetically favored and eventually turns into the global minimum, while the original minimum becomes metastable, as shown in Fig.~\ref{subfig:T_less_Tc}. The system can then transition from the metastable phase to the thermodynamically favored phase through bubble nucleation and subsequent expansion. The presence of a potential barrier separating the two phases, together with the discontinuous change of the order parameter across the transition, characterizes the transition as first order.

For a homogeneous and isotropic plasma, the nucleated regions of the new phase are expected to expand approximately spherically away from their nucleation sites. The interior of an expanding bubble is occupied by the phase with the lower free energy, while the surrounding plasma remains in the metastable phase. The interface between the two phases forms the bubble wall, across which the scalar background fields interpolate between the two phases. This picture provides the basis for the dynamical treatment of the bubble wall developed in the following sections.

We next introduce the characteristic quantities used to describe the phase transition. The first is the critical temperature $T_c$, defined as the temperature at which the free energies of the two relevant minima become degenerate,
\begin{equation}
  V_{\text{eff}}(\varphi_c, \varphi_{sc}, T_c)=V_{\text{eff}}(0,\varphi_{sc}', T_c) \,,
\end{equation}
where $(\varphi_c,\varphi_{sc})$ and $(0,\varphi_{sc}')$ denote the locations of the broken and symmetric phase minima, respectively, at $T_c$. The critical temperature characterizes the point at which the two phases are thermodynamically degenerate, but it does not by itself determine when the transition actually proceeds.	

As the Universe cools below $T_c$, the phase transition process can occur through thermal tunneling. The nucleation temperature $T_n$ is conventionally defined as the temperature at which the probability for nucleating bubbles becomes of order unity over a cosmological volume and time scale. In a simplified estimate, this condition can be expressed as
\begin{equation}
  \frac{\Gamma(T_n)}{H^4(T_n)}\approx1\,,
\end{equation}
where $\Gamma(T)$ is the bubble nucleation rate per unit volume and $H(T)$ is the Hubble expansion rate. A more precise determination can be obtained from the integrated nucleation probability, but the above criterion provides a useful estimate of the characteristic nucleation temperature used in the present analysis.

Another important quantity is the transition strength, conventionally characterized by the parameter $\alpha$. It characterizes the energy density difference between the two phases relative to the radiation energy density and can be written as
\begin{equation}
  \alpha=\frac{\Delta\rho}{\rho_R}\approx\frac{\Delta V_\text{eff}-T\frac{\partial\Delta V_\text{eff}}{\partial T}}{\frac{\pi^2g_*T^4}{30}} \,,
\end{equation}
where $g_*$ denotes the effective number of relativistic degrees of freedom at the temperature of interest. Here, the sign convention for $\Delta V_{\mathrm{eff}}$ is chosen such that it is positive when the metastable phase has a higher free energy than the thermodynamically favored phase. In the limit of strong supercooling, for which $T_n\ll T_c$, the temperature dependent contribution to the energy density difference may become subdominant, and the strength parameter can be approximated by $\alpha\approx30\Delta V_\text{eff}\big/(\pi^2g_*T^4)$.

Using the benchmark parameter point in Tab.~\ref{tab:effective_potential_benchmark}, we calculate the phase transition parameters using CosmoTransitions~\cite{Wainwright:2011kj}. The resulting critical temperature, nucleation temperature, and transition strength evaluated at $T_n$ are listed in Tab.~\ref{tab:PT_paramaters}.

\begin{table}[htbp]
\centering
\caption{Characteristic parameters of the FOEWPT for the benchmark parameter point.}
\label{tab:PT_paramaters}
\begin{tabular}{lcc}
\hline\hline
critical temperature $T_c$ &  ~~~nucleation temperature $T_n$ &  ~~~ phase transition strength $\alpha_n(T_n)$ \\ 
\hline
132.7829~[GeV]& 129.6401~[GeV] & 0.01131 \\
\hline
\end{tabular}
\end{table}

These quantities, together with the corresponding thermodynamic properties of the two phases, provide the background inputs for the subsequent analysis of bubble wall dynamics.

\section{Bubble Wall Dynamics and Determination of the Wall Velocity}
\label{sec:bubble_wall_dynamics}

We now turn to the dynamics of the bubble wall. As discussed in the previous section, a FOEWPT proceeds through the nucleation and expansion of bubbles of the thermodynamically favored phase within the metastable phase. Once a bubble approaches a steady expansion regime, its wall provides an interface between the two phases and propagates through the surrounding plasma with an approximately constant velocity. Determining this velocity requires accounting for both the macroscopic pressure difference between the two phases and the microscopic backreaction of the plasma.

\subsection{Physical setup and equation of motion}
\label{subsec:set_up_and_EOM}

\begin{figure}[htbp]
    \centering
    \includegraphics[width=0.98\textwidth]{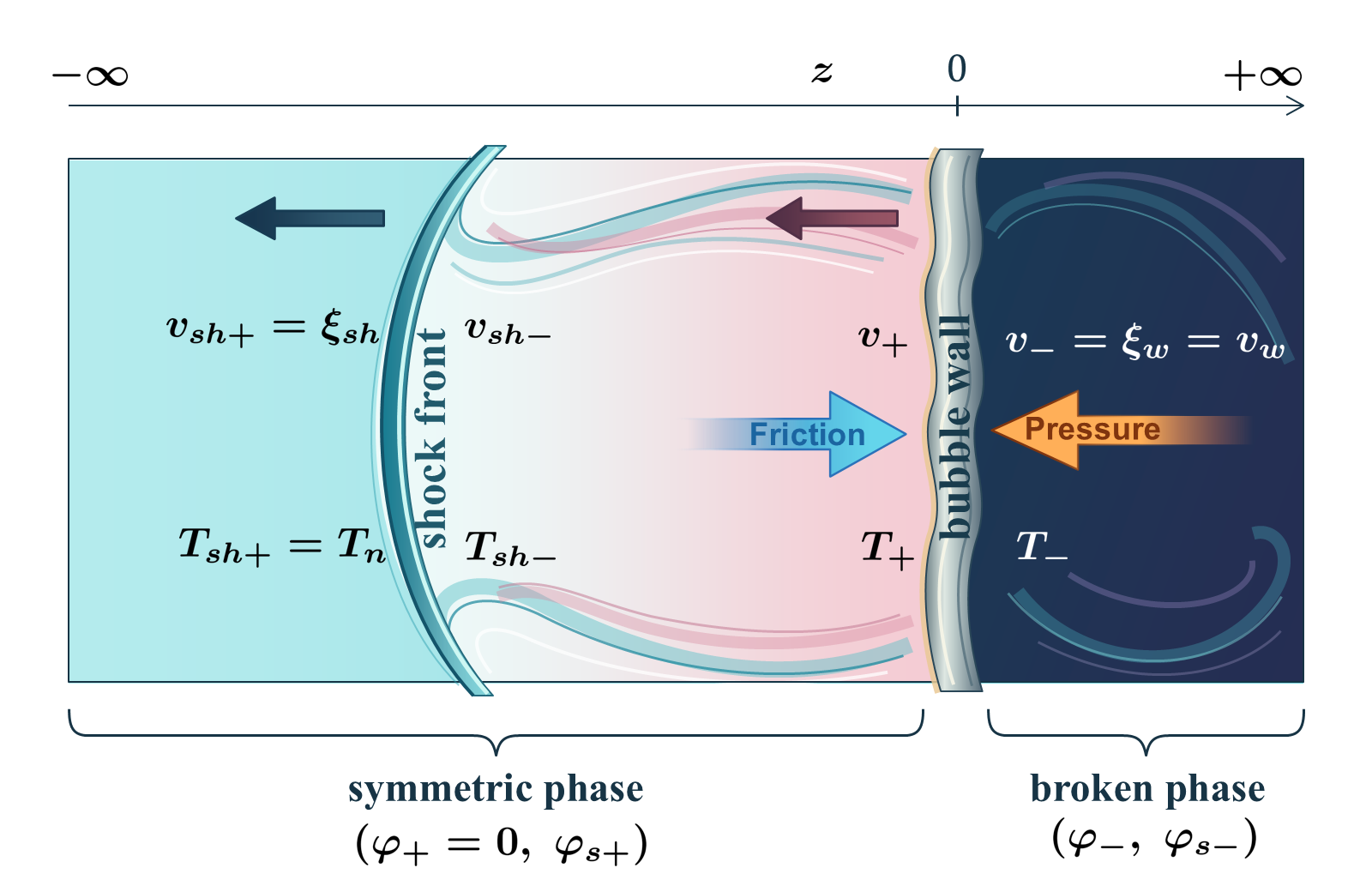}
    \caption{Schematic illustration of bubble wall dynamics in the deflagration regime. The bubble wall propagates along the negative $z$ direction, with a shock front forming ahead of it. The symmetric and broken phases are characterized by $(\varphi_+,\varphi_{s+})$ and $(\varphi_-,\varphi_{s-})$, respectively. $T_\pm$ and $v_\pm$ denote the plasma temperatures and fluid velocities on the two sides of the wall, while $T_{\mathrm{sh}\pm}$ and $v_{\mathrm{sh}\pm}$ denote the corresponding quantities across the shock front. The driving pressure and plasma friction are also indicated.}
    \label{fig:background_setup}
\end{figure}

For the isotropic and homogeneous cosmological background considered here, it is sufficient to study a locally planar segment of the bubble wall. As shown in Fig.~\ref{fig:background_setup}, we choose the wall to propagate along the negative $z$ direction and work in the wall rest frame. The broken and symmetric phases are located at $z>0$ and $z<0$, respectively, with background field values $(\varphi_-,\,\varphi_{s-})$ and $(\varphi_+=0,\,\varphi_{s+})$. 
The wall is located at $z=0$. The plasma immediately in front of and behind the wall is characterized by temperatures $T_+$ and $T_-$ and by fluid velocities $v_+$ and $v_-$, respectively, where the velocities are measured in the wall rest frame. A shock front may form ahead of the bubble wall in the deflagration regime; its hydrodynamic properties will be discussed in Sec.~\ref{subsec:Hydrodynamics}.

For a wall moving with a constant velocity \(v_w\) in the plasma frame, we introduce a coordinate comoving with the wall,
\begin{equation}
  z=z'+v_w t' \,,
\end{equation}
where $(z',t')$ denote the coordinates in the plasma frame. Under the steady state approximation, all macroscopic quantities depend only on $z$, so that
\begin{equation}
  \frac{\partial}{\partial z'}=\frac{d}{dz},
  \qquad
  \frac{\partial}{\partial t'}=v_w\frac{d}{dz} \,.
\end{equation}
In particular, the scalar field profiles interpolate between the two phases across the wall.

The physical origin of the wall acceleration and its eventual terminal velocity can be understood from the competition between the driving force and the plasma backreaction. As the background fields vary across the wall, the field dependent masses of the particles in the plasma change according to Eqs.~\eqref{eq:Higgs_mass_squre} and \eqref{eq:SM_partical_mass_squre}. Particle scattering and transport across this spatially varying mass background drive the distribution functions away from local equilibrium and generate a nonequilibrium force on the scalar fields. This contribution acts as a microscopic friction force on the wall. In contrast, the difference in free energy, or equivalently the pressure, between the two phases provides the driving force for bubble expansion. A steady state wall is therefore obtained when the net force on the wall vanishes.

To derive the corresponding EOMs, we decompose the total energy momentum tensor into the scalar field and plasma contributions $T^{\mu\nu}=T^{\mu\nu}_\text{field}+T^{\mu\nu}_\text{plasma}$. The energy momentum tensor of the scalar sector is
\begin{equation}
T^{\mu\nu}_\text{field}=\partial^{\mu}\varphi\partial^{\nu}\varphi+\partial^{\mu}\varphi_s\partial^{\nu}\varphi_s+g^{\mu\nu}\left[-\frac{1}{2}\partial^{\alpha}\varphi\partial_{\alpha}\varphi-\frac{1}{2}\partial^{\beta}\varphi_s\partial_{\beta}\varphi_s-V_{T=0}(\varphi, \varphi_s)\right] \,,
\end{equation}
where $V_{T=0}(\varphi,\varphi_s)$ denotes the zero temperature part of the scalar potential. The contribution from the plasma can be expressed kinetically in terms of the particle distribution functions,
\begin{equation}
  T^{\mu\nu}_\text{plasma}=\sum_i\int\,\frac{d^3p}{(2\pi)^3}\frac{p^{\mu}p^{\nu}}{E_i}f_i(x,p) \,,
\end{equation}
where the sum runs over the particle species included in the plasma, $f_i(x,p)$ is the corresponding phase space distribution function, and $E_i=\sqrt{p^2+m_i^2}$ is the field dependent particle energy. Imposing local energy momentum conservation $\nabla_{\mu}T^{\mu\nu}=0$ and varying with respect to the scalar background fields yields the coupled EOMs
\begin{equation}
  \begin{split}
    EOM_{\varphi}:~~~  \Box&\varphi -\frac{\partial V_{T=0}}{\partial \varphi}-\sum_i\frac{\partial m_i^2}{\partial \varphi}\int\,\frac{d^3p}{(2\pi)^3}\frac{1}{2E_i}f_i(x,p)=0 \,,\\
	  EOM_{\varphi_s}:~~~  \Box&\varphi_s -\frac{\partial V_{T=0}}{\partial \varphi_s}-\sum_i\frac{\partial m_i^2}{\partial \varphi_s}\int\,\frac{d^3p}{(2\pi)^3}\frac{1}{2E_i}f_i(x,p)=0 \,.
    \label{eq:primordial_EOM}
  \end{split}
\end{equation}

The plasma contribution to these equations depends on the full nonequilibrium distribution functions. It is therefore useful to decompose each distribution into an equilibrium part and a perturbation,
\begin{equation}
  f_i(x,p)=f_{0i}^{eq}(x,p)+\delta f_i(x,p)\,.
  \label{eq:distribution_func_part}
\end{equation}
The contribution from the equilibrium distributions reproduces the finite temperature correction to the effective potential introduced in Sec.~\ref{subsec:effective_potential}. The scalar field equations can consequently be written in terms of the full finite temperature effective potential and the nonequilibrium perturbations as
\begin{equation}
  \begin{split}
    EOM_{\varphi}:~~~  &\Box\varphi -\frac{\partial V_{\text{eff}}}{\partial \varphi}-\sum_i\frac{\partial m_i^2}{\partial \varphi}\int\,\frac{d^3p}{(2\pi)^3}\frac{1}{2E_i}\delta f_i(x,p)=0 \,,\\
	  EOM_{\varphi_s}:~~~  &\Box\varphi_s -\frac{\partial V_{\text{eff}}}{\partial \varphi_s}-\sum_i\frac{\partial m_i^2}{\partial \varphi_s}\int\,\frac{d^3p}{(2\pi)^3}\frac{1}{2E_i}\delta f_i(x,p)=0 \,.
    \label{eq:mid_EOM}
  \end{split}
\end{equation}
This form separates the two effects that determine the wall dynamics. The effective potential accounts for the equilibrium thermodynamic driving force, while the $\delta f_i$ terms encode the microscopic nonequilibrium response of the plasma and hence the friction experienced by the wall.

Finally, transforming to the wall rest frame and using the steady state condition, the equations reduce to ordinary differential equations in $z$,
\begin{equation}
  \begin{split}
    EOM_{\varphi}:~~~ &\left(1-v_w^2\right)\frac{d^2\varphi}{dz^2}-\frac{\partial V_{\text{eff}}}{\partial \varphi}-\sum_i\frac{\partial m_i^2}{\partial \varphi}\int\,\frac{d^3p}{(2\pi)^3}\frac{1}{2E_i}\delta f_i(z,p)=0  \,,\\
	  EOM_{\varphi_s}:~~~ &\left(1-v_w^2\right)\frac{d^2\varphi_s}{dz^2}-\frac{\partial V_{\text{eff}}}{\partial \varphi_s}-\sum_i\frac{\partial m_i^2}{\partial \varphi_s}\int\,\frac{d^3p}{(2\pi)^3}\frac{1}{2E_i}\delta f_i(z,p)=0	\,,
    \label{eq:final_EOM}
  \end{split}
\end{equation}
where the factor $(1-v_w^2)$ comes from the steady-state travelling wall ansatz and the coordinate transformation between the plasma frame and the wall frame. These equations provide the starting point for the determination of the wall velocity. The remaining task is to compute the plasma state on both sides of the wall and, in particular, the nonequilibrium perturbations $\delta f_i$ induced by the spatially varying scalar background.

\subsection{Hydrodynamics of the plasma}
\label{subsec:Hydrodynamics}

The scalar field equations derived in the previous section determine the response of the phase transition fields to the driving force and to the nonequilibrium plasma friction. At the same time, the propagation of the bubble wall perturbs the surrounding plasma and induces nontrivial profiles of the fluid velocity and temperature. These macroscopic plasma effects must be taken into account when determining the thermodynamic state near the wall. We therefore describe the plasma within the hydrodynamic approximation and impose energy momentum conservation across the bubble wall and the shock front.

We approximate the plasma as a relativistic ideal fluid with energy momentum tensor
\begin{equation}
	T^{\mu\nu}_{f}=(e_f+p_f)u^{\mu}u^{\nu}+p_f g^{\mu\nu}=w_f u^{\mu}u^{\nu}+p_f g^{\mu\nu} \,,
  \label{eq:energy_momentum_tensor}
\end{equation}
where $e_f$, $p_f$, and $w_f=e_f+p_f$ denote the energy density, pressure, and enthalpy density, respectively, and $u^\mu=\gamma(v)(1,\boldsymbol{v})$ is the fluid four-velocity with $\gamma(v)=(1-v^2)^{-1/2}$.

For a steadily expanding spherical bubble, the hydrodynamic profiles are self-similar to a good approximation and depend on the radial coordinate $r$ and time $t$ only through $\xi=r/t$. Using $\nabla_\mu T_f^{\mu\nu}=0$ in spherical coordinates and rewriting the resulting equations in terms of $\xi$, one obtains the energy and momentum conservation equations
\begin{gather}
  (v-\xi)\partial_{\xi}(w_f\gamma^2)+\xi\partial_{\xi}p_f+w_f\gamma^2\partial_{\xi}v+\frac{2w_f\gamma^2v}{\xi}=0 \,,
  \label{eq:energy_conserve}\\
  v(v-\xi)\partial_{\xi}(w_f\gamma^2)+w_f\gamma^2(2v-\xi)\partial_{\xi}v+\partial_{\xi}p_f+\frac{2w_f\gamma^2v^2}{\xi}=0 \,.
  \label{eq:momenten_conserve}
\end{gather}

Combining these equations gives the standard hydrodynamic equations for the fluid velocity and energy density,
\begin{gather}
	(1-v\xi)\frac{\partial_{\xi}p_f}{w_f}=\gamma^2(\xi-v)\partial_{\xi}v \,, \label{eq:hydro_eq1} \\
	(\xi-v)\frac{\partial_{\xi}e_f}{w_f}=\frac{2v}{\xi}+[1-\gamma^2v(\xi-v)]\partial_{\xi}v \,,   \label{eq:hydro_eq2}
\end{gather}
where the local sound speed is defined by $c_s^2\equiv(dp_f/dT)\big/(de_f/dT)$ and $\mu(\xi,v)=(\xi-v)\big/(1-\xi v)$ denotes the fluid velocity in the local plasma frame obtained by Lorentz transforming the fluid velocity $v$ from the self-similar frame. The coupled hydrodynamic equations can be reduced to a single differential equation for the velocity profile,
\begin{equation}
	\frac{2v}{\xi}=\gamma^2(1-v\xi)\left[\frac{\mu^2}{c_s^2}-1\right]\partial_{\xi}v \,,
	\label{eq:v_profile}
\end{equation}
while the temperature profile follows from the enthalpy relation $w_f(T)\equiv T\frac{\partial p_f}{\partial T}=T\partial_{\xi}p_f(\partial_{\xi}T)^{-1}$ leading to
\begin{equation}
	\frac{\partial_{\xi}T}{T}=\gamma^2\mu\partial_{\xi}v \,.
	\label{eq:T_profile}
\end{equation}

These two equations determine the macroscopic plasma profiles once suitable boundary conditions are specified. The boundary conditions at the phase transition front follow from energy momentum conservation across the wall. In the wall rest frame, the fluid variables on the two sides satisfy
\begin{equation}
	v_{+}v_{-}=\frac{p_{+}-p_{-}}{e_{+}-e_{-}},~~~ \frac{v_{+}}{v_{-}}=\frac{e_{-}+p_{+}}{e_{+}+p_{-}} \,,
	\label{eq:matching_condition}
\end{equation}
where the subscripts $+$ and $-$ refer to the plasma immediately in front of and behind the wall, respectively. To obtain an analytic relation between $v_+$ and $v_-$, we adopt the bag equation of state as an approximation for the two phases,
\begin{equation}
  \begin{split}
    p_{+}=&\frac{1}{3}a_{+}T_{+}^4-\epsilon_{+},~~~e_{+}=a_{+}T_{+}^4+\epsilon_{+},~~~~\epsilon_{+}\equiv V_{T=0}(0,~\varphi_{s+}) \,,\\
	  p_{-}=&\frac{1}{3}a_{-}T_{-}^4-\epsilon_{-},~~~e_{-}=a_{-}T_{-}^4+\epsilon_{-},~~~~\epsilon_{-}\equiv V_{T=0}(\varphi_-,~\varphi_{s-}) \,.
    \label{eq:bag_model_EOS}
  \end{split}
\end{equation}
Here, $a_\pm=\pi^2g_{\ast,\pm}/30$ are the radiation density coefficients, while $\epsilon_\pm$ denote the vacuum energy contributions in the two phases. Under the approximation $T_+\simeq T_-$, which is adequate for the relatively weakly supercooled transitions considered here, the matching conditions can be reduced to
\begin{equation}
	v_{+}=\frac{1}{1+\alpha_{+}}\left[\left(\frac{v_{-}}{2}+\frac{1}{6v_{-}}\right)\pm \sqrt{\left(\frac{v_{-}}{2}+\frac{1}{6v_{-}}\right)^2+\alpha_{+}^2+\frac{2}{3}\alpha_{+}-\frac{1}{3}}\right] \,,
	\label{eq:v_front}
\end{equation}
where $\alpha_{+}\equiv \frac{\epsilon_{+}-\epsilon_{-}}{a_{+}T_{+}^4}$ characterizes the transition strength evaluated using the plasma temperature immediately in front of the wall.

The hydrodynamic matching relation admits several branches corresponding to different modes of bubble expansion. The detonation branch describes a supersonic wall with an undisturbed plasma in front of the wall, while the deflagration branch corresponds to a subsonic wall preceded by a shock front. A hybrid branch can also occur, with a supersonic fluid velocity in front of the wall and a subsonic fluid velocity behind it. In the present work, we focus on the deflagration regime and therefore select the corresponding branch of the matching relation.

For clarity, we distinguish three different velocity frames throughout the calculation. The quantities $v_\pm$ denote the fluid velocities measured in the wall rest frame, $v_{\mathrm{sh}\pm}$ denote the fluid velocities in the shock front rest frame, and $\tilde v_\pm$ denote the corresponding velocities in the plasma frame. The velocities in different frames are related by the relativistic velocity-addition formula,
\begin{equation}
	\tilde{v}_{\pm}=\frac{v_{w}-v_{\pm}}{1-v_{w}v_{\pm}},~~~\tilde{v}_{sh\pm}=\frac{v_{sh}-v_{sh\pm}}{1-v_{sh}v_{sh\pm}} \,.
\end{equation}

For a deflagration solution, the plasma behind the wall is at rest far from the bubble, implying $\tilde{v}_{-}=0~\rightarrow~v_w=v_{-}=\xi_w$. Similarly, the plasma ahead of the shock front is taken to be at rest in the plasma frame, so that $\tilde{v}_{sh+}=0~\rightarrow~v_{sh}=v_{sh+}=\xi_{sh}$. The velocity profile is therefore integrated from the wall position $\xi_w=v_w$ to the shock front at $\xi_{\mathrm{sh}}$. At the wall, the boundary condition is determined by the wall matching relation,
\begin{equation}
	\tilde{v}_{+}=\frac{\xi_w-v_{+}}{1-\xi_wv_{+}}=v(\xi_w) \,,
	\label{eq:v_bond_condition}
\end{equation}
while the shock position is fixed by the matching conditions across the shock front,
\begin{equation}
	\tilde{v}_{sh-}=\frac{\xi_{sh}-v_{sh-}}{1-\xi_{sh}v_{sh-}}=v(\xi_{sh}) \,.
\end{equation} 
Together with the relativistic velocity transformation, this yields the condition
\begin{equation}
	\mu(\xi_{sh}, v(\xi_{sh}))=v_{sh-}~\rightarrow~\mu(\xi_{sh}, v(\xi_{sh}))\xi_{sh}=\frac{1}{3}=c_s^2 \,.
	\label{eq:shock_bond_condition}
\end{equation}

For a given trial value of the wall velocity $v_w$ and transition strength $\alpha_+$, the hydrodynamic velocity profile can therefore be obtained by integrating the velocity equation from $\xi_w$ toward larger $\xi$ until the shock front condition is reached. The temperature profile is then obtained by integrating the corresponding temperature equation. In this way, both $v(\xi)$ and $T(\xi)$ are determined throughout the region between the shock front and the bubble wall.

The temperature immediately in front of the wall, $T_+$, is related to the nucleation temperature $T_n$ through the heating induced by the shock front and the plasma flow. Taking the temperature ahead of the shock to be $T_{\mathrm{sh}+}=T_n$, the shock matching conditions give
\begin{equation}
	\frac{T_{sh-}^4}{T_n^4}=\frac{T_{sh-}^4}{T_{sh+}^4}=\frac{3(1-v_{sh-}^2)}{9v_{sh-}^2-1} \,.
\end{equation}
Combining this relation with the integrated temperature profile yields
\begin{equation}
	\frac{T_{+}}{T_n}=\left[\frac{3(1-v_{sh-}^2)}{9v_{sh-}^2-1}\right]^{1/4}\exp\left[\int_{\xi_{sh}}^{\xi_w}\,d\xi\,\frac{2c_s^2v(\xi-v)}{\xi[(\xi-v)^2-c_s^2(1-v\xi)^2]}\right] \,.
\end{equation}
The temperature behind the wall, $T_-$, is then obtained from the enthalpy matching condition, ${a_+T_+^4}/{a_-T_-^4}={w_+}/{w_-}$.

An important point is that $\alpha_+$ cannot in general be identified directly with the transition strength $\alpha_n$ evaluated at the nucleation temperature. In the numerical calculation, we therefore use $\alpha_n$ only as an initial estimate for $\alpha_+$. For a given trial $v_w$, the hydrodynamic solution first determines $T_+$; the transition strength is then recalculated at this temperature and used as the input for the next iteration. The procedure is repeated until $T_+$ and $\alpha_+$ converge to a self-consistent solution.

\begin{figure}[htbp]
    \centering
    \includegraphics[width=0.98\textwidth]{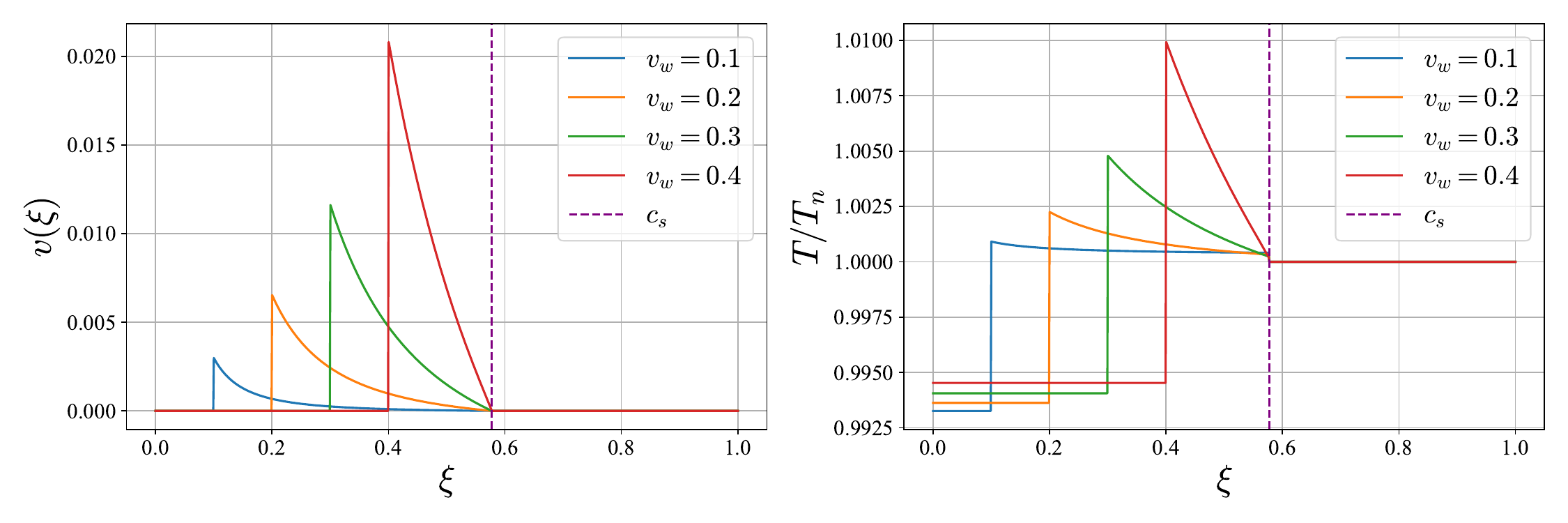}
    \caption{Fluid velocity and temperature profiles in the deflagration regime for a benchmark transition with $\alpha_n\simeq0.01$. The profiles are shown for several assumed bubble wall velocities; the dashed line indicates the sound speed.}
    \label{fig:deflagration_profile}
\end{figure}

Figure~\ref{fig:deflagration_profile} shows representative velocity and temperature profiles obtained in the deflagration regime for a benchmark transition with $\alpha_n\simeq0.01$ and several assumed values of $v_w$.

The self-similar coordinate decreases from the unperturbed plasma far ahead of the bubble toward the bubble interior. The shock front heats and accelerates the plasma, producing increasing temperature and fluid velocity toward the wall. The corresponding quantities undergo a discontinuous jump across the bubble wall in the hydrodynamic treatment, since the wall thickness is not resolved explicitly. The hydrodynamic approximation therefore determines the macroscopic plasma state on both sides of the wall without resolving the microscopic wall profile itself. These results provide the background plasma conditions required for the microscopic transport calculation in the next section.

\subsection{Microscopic particle transport}
\label{subsec:Microscopic_particle_transport}

The hydrodynamic treatment introduced above describes the plasma as a locally equilibrated fluid and captures its macroscopic response to the propagating bubble wall. Such a description, however, does not fully account for the nonequilibrium particle transport generated by the spatially varying scalar background. As particles cross the wall, their field dependent masses change, driving their distribution functions away from equilibrium. These nonequilibrium perturbations provide the microscopic origin of the friction acting on the bubble wall and therefore play a central role in its velocity determination.

\subsubsection{Boltzmann equations and perturbations}
\label{subsubsec:Boltzmann_eq}

In the present CxSM benchmark scenarios, we explicitly track the particle species whose mass variations and interactions give the dominant contribution to the nonequilibrium dynamics. Since the two CP-even scalar mass eigenstates $h_1$ and $h_2$ are approximately mass degenerate for the parameter choices considered here, we treat them as a single effective species, denoted by $H$, in the subsequent transport calculation. The additional scalar $\chi$ is not subject to this degeneracy and is therefore treated as a separate species. Together with the top quark and the electroweak gauge bosons, these constitute the heavy species included explicitly in the transport equations. For simplicity, the $W$ and $Z$ bosons are treated as a single effective gauge boson species, denoted by $W$. The remaining light degrees of freedom are assumed to remain close to kinetic equilibrium and are included as a background plasma.

We describe the particle distributions using the semiclassical Boltzmann equation. This treatment is applicable when the scalar background varies slowly on microscopic length scales, so that the WKB approximation is valid.
\begin{equation}
	\frac{d}{dt}f=\left(\frac{\partial}{\partial t}+\dot{z}\frac{\partial}{\partial z}+\dot{p_z}\frac{\partial}{\partial p_z}\right)f=-C[f] \,,
	\label{eq:boltzmanneq}
\end{equation}
where $C[f]$ denotes the collision operator and $\dot{z}=p_z\big/E$. Following the decomposition introduced in Sec.~\ref{subsec:set_up_and_EOM}, the distribution function of each particle species is written as an equilibrium contribution plus a small perturbation. To linear order, we parametrize the perturbation in terms of a chemical potential, a temperature perturbation, and a fluid-velocity perturbation,
\begin{equation}
    f=\frac{1}{e^{(E+\delta)/T}\pm1}
    =f_0+\frac{\partial f_0}{\partial E/T}\left(-\delta\mu-\frac{E}{T}\delta T-p_z\delta v\right)\cdot\frac{1}{T}
    \equiv f_0+\delta f \,.
    \label{eq:distribution_func}
\end{equation}

Since the equilibrium chemical potential vanishes for the particle species considered here, the chemical potential perturbation itself is denoted by $\mu_i$. We further allow the heavy species to interact with a common background plasma characterized by the perturbations $\mu_{\rm bg}$, $\delta T_{\rm bg}$, and $\delta v_{\rm bg}$. The latter describe the collective response of the light degrees of freedom that are not evolved as independent particle species.

Substituting the linearized distribution functions into the Boltzmann equation and retaining terms to first order in the perturbations gives a coupled set of differential equations for the perturbation variables. The derivatives acting on the equilibrium distribution generate source terms proportional to the spatial variation of the field dependent masses, while the collision operator drives the system toward local equilibrium. After neglecting higher-order derivatives of the equilibrium distribution, the linearized Boltzmann equation can be written as
\begin{equation}
	\begin{split}
		(-f_0)'\big(\frac{p_z}{E}\left[\partial_z(\mu+\mu_{bg})+\frac{E}{T}\partial_z(\delta T+\delta T_{bg})+p_z\partial_z(\delta v+\delta v_{bg})\right]+&\partial_t\mu\\
		+\frac{E}{T}\partial_t(\delta T+\delta T_{bg})+p_z\partial_t(\delta v+\delta v_{bg})\big)+TC[f]=&(-f_0)'\frac{\partial_t(m^2)}{2E} \,,
		\label{eq:full_Boltzmanneq}
	\end{split}
\end{equation}
where $(f_0)'={\partial f_0}/{\partial (E/T)}$. The resulting momentum dependence prevents the full Boltzmann equation from being solved analytically. We therefore employ a moment expansion and project the transport equation onto three independent momentum moments $1,\, E,\,$ and $p_z$. This yields three coupled differential equations for each heavy particle species. After transforming to the wall rest frame, the equations can be reduced to
\begin{equation}
	\begin{split}
		v_wc_2^i(\mu_i'+\mu_{bg}')+v_wc_3^i(\delta T_i'+\delta T_{bg}')+\frac{c_3^iT}{3}(\delta v_i'+\delta v_{bg}')+\mu_i\Gamma_{\mu1,i}+\delta T_i\Gamma_{T1,i}&=\frac{v_wc_1^i}{2T}(m_i^2)'\,,\\
		v_wc_3^i(\mu_i'+\mu_{bg}')+v_wc_4^i(\delta T_i'+\delta T_{bg}')+\frac{c_4^iT}{3}(\delta v_i'+\delta v_{bg}')+\mu_i\Gamma_{\mu2,i}+\delta T_i\Gamma_{T2,i}&=\frac{v_wc_2^i}{2T}(m_i^2)'\,,\\
		\frac{c_3^i}{3}(\mu_i'+\mu_{bg}')+\frac{c_4^i}{3}(\delta T_i'+\delta T_{bg}')+\frac{v_wc_4^iT}{3}(\delta v_i'+\delta v_{bg}')+T\delta v_i\Gamma_{v,i}&=0\,,
	\end{split}
\end{equation}
where a prime denotes differentiation with respect to the wall frame coordinate $z$. The coefficients entering the moment equations are defined by momentum integrals over the equilibrium distribution,
\begin{equation}
	c_j^{b/f}=\int\,\frac{E^{j-2}}{T^{j+1}}(-f_0')\frac{d^3p}{(2\pi)^3} \,,
  \label{eq:c_b_f}
\end{equation}
and the moments of the collision operator are parametrized as
\begin{equation}
	\begin{split}
		\int\,\frac{d^3p}{(2\pi)^3T^2}C[f_i]&=\mu_i\Gamma_{\mu1,i}+\delta T_i\Gamma_{T1,i} \,\\
		\int\,\frac{d^3p}{(2\pi)^3T^3}EC[f_i]&=\mu_i\Gamma_{\mu2,i}+\delta T_i\Gamma_{T2,i} \,\\
		\int\,\frac{d^3p}{(2\pi)^3T^3}p_zC[f_i]&=T\delta v_i\Gamma_{v,i} \,.
		\label{eq:collision_term_integrate}
	\end{split}
\end{equation}
In the present calculation, the index $i$ runs over the four effective heavy species, $i=H,~\chi,~W,~t$. The collision terms account for interactions among the heavy species and with the light particle background. Their explicit evaluation is given in the following subsection.

The light degrees of freedom are treated collectively because their interaction rates are sufficiently large to maintain approximate chemical equilibrium. We therefore set $\mu_{\rm bg}=0$ and evolve only their common temperature and velocity perturbations. Taking the sum over the heavy species and imposing energy and momentum conservation between the heavy and background components gives the corresponding background equations, \begin{equation}
	\begin{split}
		\mu_{bg}&=0 \,,\\
		\tilde{c}_4\left(v_w\delta T_{bg}'+\frac{\delta v_{bg}'}{3}T\right)&=\sum_iN_i(\mu_i\Gamma_{\mu2,i}+\delta T_i\Gamma_{T2,i}) \,,\\
		\frac{\tilde{c}_4}{3}(\delta T_{bg}'+v_wT\delta v_{bg}')&=\sum_iN_iT\delta v_i\Gamma_{v,i} \,,
	\end{split}
\end{equation}
where $\tilde c_4$ denotes the total heat capacity coefficient of the background plasma. For the particle content adopted here, $\tilde{c}_4=78c_4^f+18c_4^b\approx 113.5$.

The opposite signs of the collision contributions in the background equations follow from momentum and energy conservation: the momentum and energy transferred from the heavy species are gained by the background plasma and vice versa. Solving the background equations gives the derivatives of the temperature and velocity perturbations in terms of those of the heavy species,
\begin{equation}
	\begin{split}
		\delta v_{bg}' &=\frac{9v_w\sum_iN_iT\delta v_i\Gamma_{v,i}-3\sum_iN_i(\mu_i\Gamma_{\mu2,i}+\delta T_i\Gamma_{T2,i})}{\tilde{c}_4T(3v_w^2-1)} \,,\\
		\delta T_{bg}'&=\frac{3v_w\sum_iN_i(\mu_i\Gamma_{\mu2,i}+\delta T_i\Gamma_{T2,i})-3\sum_iN_iT\delta v_i\Gamma_{v,i}}{\tilde{c}_4(3v_w^2-1)} \,.
		\label{eq:bg_perturbation}
	\end{split}
\end{equation}
The transport problem is therefore reduced to a coupled first-order system for the perturbations of the heavy species together with the collective background response. These perturbations determine the nonequilibrium contribution to the scalar field EOMs and hence the microscopic friction acting on the bubble wall. The collision operator and its numerical evaluation are discussed next.

\subsubsection{Collision terms}
\label{subsubsec:collision_terms}

The collision operator introduced above accounts for the interactions that drive the particle distributions toward local equilibrium and therefore determines the relaxation rates entering the transport equations. In the present calculation, we retain the dominant scattering processes between the explicitly evolved heavy species and the light background plasma, while neglecting subleading channels and higher order corrections.

For a generic $2\rightarrow2$ process, $i+j\rightarrow m+n$, the collision term for particle species $i$ can be written in the standard form 
\begin{equation}
	C[f_i]=\frac{1}{2N_i}\sum\frac{1}{2E_p}\int\frac{d^3kd^3p'd^3k'}{(2\pi)^9 2E_k 2E_{p'} 2E_{k'}}|\mathcal{M}|^2(2\pi)^4\delta^{(4)}(p+k-p'-k')\mathcal{P}[f_i(p)] \,,
  \label{eq:collision_term}
\end{equation}
where $p$, $k$, $p'$, and $k'$ denote the four-momenta of the incoming and outgoing particles, respectively, and $|\mathcal M|^2$ is the corresponding spin and color summed matrix element. The statistical factor is given by
\begin{equation}
	\mathcal{P}[f_i(p)]=f_i(p)f_j(k)(1\mp f_m(p'))(1\mp f_n(k'))-f_m(p')f_n(k')(1\mp f_i(p))(1\mp f_j(k)) \,,
  \label{eq:statistic_factor}
\end{equation}
with the upper and lower signs corresponding to fermionic and bosonic statistics, respectively.

To obtain the linearized transport equations, we expand the distribution functions to first order around equilibrium. For notational convenience, we write $f_i\equiv f_{0i}-f_{0i}'\cdot{\delta_i}/{T}$, where $\delta_i$ collects the perturbations associated with the chemical potential, temperature, and fluid velocity of particle species $i$. Substituting the linearized distributions into the statistical factor and retaining only terms up to first order in the perturbations gives
\begin{equation}
    \mathcal{P}[f_i(p)]
    =\frac{1}{T}(\delta_i+\delta_j-\delta_m-\delta_n)f_{0i}f_{0j}(1\mp f_{0m})(1\mp f_{0n}) \,.
\end{equation}

Using energy conservation in the scattering process and neglecting terms of second and higher order in the perturbations, the collision term reduces to the linear form
\begin{equation}
	\begin{split}
		C[f_i]=\frac{1}{2N_i}\sum&\frac{1}{2E_p}\int\frac{d^3kd^3p'd^3k'}{(2\pi)^9 2E_k 2E_{p'} 2E_{k'}}|\mathcal{M}|^2(2\pi)^4\delta^{(4)}(p+k-p'-k') \\
		&\times\frac{1}{T}(\delta_i+\delta_j-\delta_m-\delta_n)f_{0i}f_{0j}(1\mp f_{0m})(1\mp f_{0n}) \,.
    \label{eq:collision_term_final}
	\end{split}
\end{equation}
This expression makes explicit that the collision operator vanishes in thermal equilibrium and is proportional, at leading order, to the difference between the perturbations of the incoming and outgoing states. 

The collision coefficients entering the moment equations are obtained by inserting the linearized collision term into the three momentum moments defined in Eq.~\eqref{eq:collision_term_integrate}. We sum over the relevant scattering channels for each effective heavy species, $i=H,\chi,W,t,$ and evaluate the resulting multidimensional phase space integrals numerically. The explicit scattering matrix elements and the reduction of the phase space integrals to a numerically tractable form are collected in Apps.~\ref{app:scattering_matrix} and \ref{app:phase_space_integration}, respectively.

For the benchmark point listed in Tab.~\ref{tab:effective_potential_benchmark}, the resulting collision coefficients are of order $10^{-3}T$ for the top quark and gauge boson sectors, while the scalar sectors exhibit comparatively larger relaxation coefficients. The numerical values for this benchmark are listed in App.~\ref{app:numerical_collsion}; the collision coefficients are recalculated for each parameter point in the subsequent scan. In the parameter space analysis presented in Sec.~\ref{sec:numerical_results}, the collision coefficients are recalculated consistently for each benchmark point because the relevant particle masses and interaction rates vary with the underlying CxSM parameters.

The collision coefficients therefore provide the microscopic relaxation input required to solve the coupled transport equations. Once evaluated for a given parameter point, they are combined with the hydrodynamic background and the scalar field profiles to determine the nonequilibrium perturbations and the resulting microscopic friction on the bubble wall.

\subsubsection{Solution of the transport equations}

With the collision coefficients determined as described above, the coupled transport equations can be written in a compact matrix form,
\begin{equation}
	\hat{A}\delta'+\hat{\Gamma}\delta=\hat{\Sigma} \,,
    \label{eq:transport}
\end{equation}
where the perturbation vector contains the chemical potential, temperature, and velocity perturbations of the four effective heavy species,
\begin{equation}
	\delta=(\mu_t,~\delta T_t,~T\delta v_t,~\mu_W,~\delta T_W,~T\delta v_W,~\mu_H,~\delta T_H,~T\delta v_H,~\mu_\chi,~\delta T_\chi,~T\delta v_\chi)^T \,,
\end{equation}
and the source vector is determined by the spatial variation of the field dependent masses,
\begin{equation}
	\hat{\Sigma}=\frac{v_w}{2T}(\hat{\Sigma_t},\,\hat{\Sigma_W},\,\hat{\Sigma_H},\,\hat{\Sigma_\chi})^T,~~~\hat{\Sigma_i}=(c_1^i(\bar{m}^2_i)',\,c_2^i(\bar{m}^2_i)',\,0) \,.
\end{equation}

The matrix $\hat A$ contains the coefficients of the derivative terms and is block diagonal in the heavy species basis,
\begin{equation}
	\hat{A}=\begin{pmatrix}
			\hat{A}_t & 0 & 0 & 0 \\
			0 & \hat{A}_W & 0 & 0 \\
			0 & 0 & \hat{A}_H & 0 \\
			0 & 0 & 0 & \hat{A}_\chi
	\end{pmatrix}
	,~~~\hat{A}_i=\begin{pmatrix}
		v_wc_2^i & v_wc_3^i & \frac{c_3^i}{3} \\
		v_wc_3^i & v_wc_4^i & \frac{c_4^i}{3} \\
		\frac{c_3^i}{3} & \frac{c_4^i}{3} & \frac{v_wc_4^i}{3}
	\end{pmatrix}
	\,,
\end{equation}
while the nondifferential terms are collected in
\begin{gather}
\hat{\Gamma}=\hat{\Gamma}_0+\frac{1}{\tilde{c_4}}\hat{\Gamma}_{bg} \,,\\
\hat{\Gamma}_0=\begin{pmatrix}
\Gamma_t & 0 & 0 & 0 \\
0 & \Gamma_W & 0 & 0 \\
0 & 0 & \Gamma_H & 0 \\
0 & 0 & 0 & \Gamma_\chi
\end{pmatrix}
,~~~\Gamma_i=\begin{pmatrix}
    \Gamma_{\mu1,i} & \Gamma_{T1,i} & 0 \\
    \Gamma_{\mu2,i} & \Gamma_{T2,i} & 0 \\
    0 & 0 & \Gamma_{v,i}
\end{pmatrix}
\,,\\
\hat{\Gamma}_{bg}=\begin{pmatrix}
    \Gamma_{tt} & \Gamma_{tW} & \Gamma_{tH} & \Gamma_{t\chi} \\
    \Gamma_{Wt} & \Gamma_{WW} & \Gamma_{WH} & \Gamma_{W\chi} \\
    \Gamma_{Ht} & \Gamma_{HW} & \Gamma_{HH} & \Gamma_{H\chi} \\
    \Gamma_{\chi t} & \Gamma_{\chi W} & \Gamma_{\chi H} & \Gamma_{\chi\chi}
\end{pmatrix}
,~~~\Gamma_{ij}=N_j\begin{pmatrix}
    c_3^i\Gamma_{\mu2,j} & c_3^i\Gamma_{T2,j} & 0 \\
    c_4^i\Gamma_{\mu2,j} & c_4^i\Gamma_{T2,j} & 0 \\
    0 & 0 & c_4^i\Gamma_{v,j}
\end{pmatrix}
\,.
\end{gather}
The background contribution accounts for the collective response of the light plasma degrees of freedom derived in the previous subsection.

The transport equations can therefore be recast as a first-order linear system,
\begin{equation}
	\delta'+\hat{A}^{-1}\hat{\Gamma}\delta=\hat{A}^{-1}\hat{\Sigma} \,.
\end{equation}
To obtain an analytic representation of the solution, we diagonalize the matrix $\hat A^{-1}\hat\Gamma$,
\begin{equation}
	(\hat{A}^{-1}\hat{\Gamma})_{ij}\Lambda_{jk}=\Lambda_{ik}\rho_k \,,
\end{equation}
where $\rho_k$ are the eigenvalues and the columns of $\Lambda$ are the corresponding eigenvectors. Introducing the transformed variables $\eta=\Lambda^{-1}\delta$, the coupled system is reduced to a set of independent first-order equations for the individual eigenmodes,
\begin{equation}
    \left(\frac{d}{dz}+\rho_k\right)\eta_k=[\Lambda^{-1}\hat{A}^{-1}\hat{\Sigma}]_k \,.
\end{equation}
In the approximation adopted here, $\hat A^{-1}\hat\Gamma$ is treated as independent of $z$ within the transport calculation, so that the eigenvector matrix $\Lambda$ is constant across the wall. Imposing the physical boundary condition $\delta(z\rightarrow\pm\infty)=0$, the Green's function for the $k$-th eigenmode can be written as
\begin{equation}
	\mathcal{G}_k(z,y)=\text{sgn}(\rho_k)e^{-\rho_k(z-y)}\Theta(\text{sgn}(\rho_k)(z-y)) \,.
\end{equation}
The perturbations are then obtained by convoluting the Green's functions with the source terms,
\begin{equation}
	\delta_i(z)=\Lambda_{ik}\int_{-\infty}^{+\infty}\,dy\,\mathcal{G}_k(z,y)[\Lambda^{-1}\hat{A}^{-1}\hat{\Sigma}]_k \,.
\end{equation}
This representation makes clear that the nonequilibrium perturbations at a given position are determined by the spatially varying source generated by the scalar field background, together with the relaxation properties encoded in the collision matrix.

The source vector depends explicitly on the derivatives of the field dependent masses and therefore requires a specification of the scalar field profile across the wall. We adopt the standard smooth wall parametrization
\begin{equation}
	\begin{split}
		\varphi(z)&=\frac{\varphi_-}{2}\left(1+\tanh\frac{z}{L_w}\right) \,,\\ 
		\varphi_s(z)&=\frac{\varphi_{s-}-\varphi_{s+}}{2}\left(1+\tanh\frac{z}{L_w}\right)+\varphi_{s+} \,,
		\label{eq:wall_profile}
	\end{split}
\end{equation}
where $L_w$ characterizes the width of the bubble wall. The $\tanh$ ansatz provides a convenient approximation to the smooth interpolation between the symmetric and broken phases and allows the source terms $(\bar m_i^2)'$ to be evaluated analytically.

\begin{figure}[htbp]
    \centering
    \includegraphics[width=0.98\textwidth]{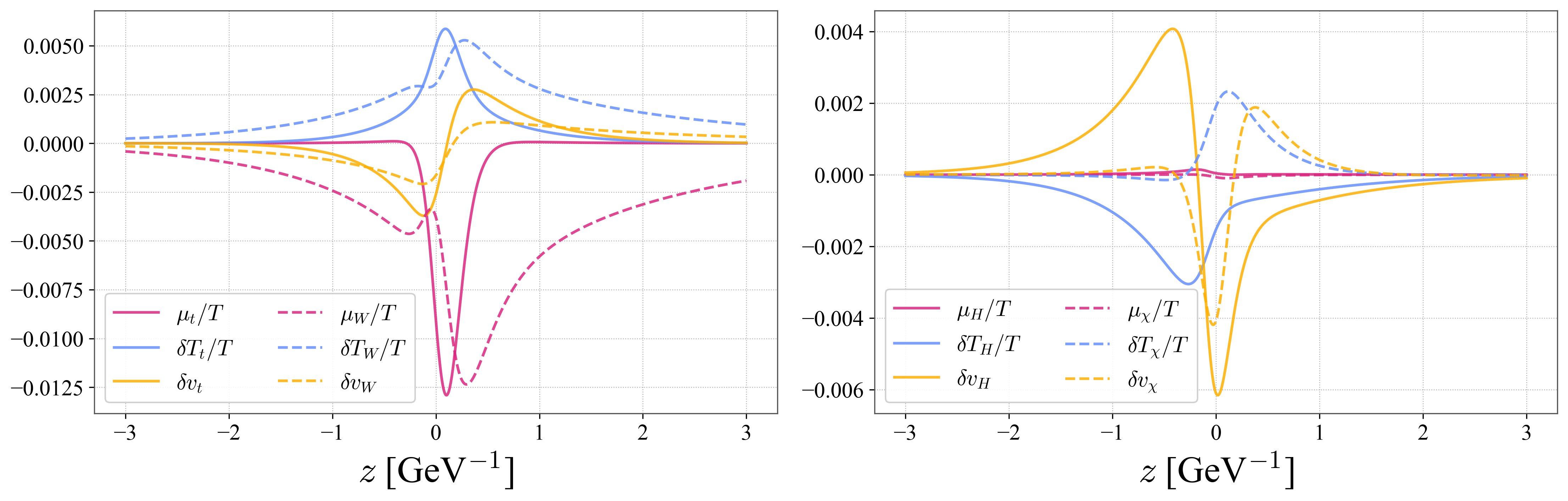}
    \caption{Profiles of the perturbations $\mu_i(z)/T$, $\delta T_i(z)/T$, and $\delta v_i(z)$ for the $t$, $W$, $H$, and $\chi$ particles at $v_w=0.1$ and $L_w=0.2~\mathrm{GeV}^{-1}$. The left panel shows the perturbations of the $t$ (solid) and $W$ (dashed) particles, while the right panel shows those of the $H$ (solid) and $\chi$ (dashed) particles.}
    \label{fig:perturbations}
\end{figure}

Using the wall profiles together with the collision coefficients and hydrodynamic background obtained above, we solve the transport equations numerically for the perturbations of the heavy species. A representative result for the benchmark parameter point is shown in Fig.~\ref{fig:perturbations}.

Figure~\ref{fig:perturbations} describes the spatial profiles of the perturbations across the bubble wall and illustrates their localization around the region where the scalar background varies most rapidly. In particular, the chemical potential and temperature perturbations are generated by the changing particle masses and are subsequently modified by collisional relaxation. The relative magnitude and sign of the perturbations depend on the particle species and on the competition between the wall induced source and the corresponding relaxation rates. These nonequilibrium perturbations enter directly into the scalar field EOMs through the friction terms and therefore provide the microscopic input for the determination of the steady state wall velocity in the next subsection.

The benchmark point result shown here is intended to illustrate the structure of the transport solution. The perturbations are recalculated for each benchmark parameter set considered in Sec.~\ref{sec:numerical_results}, using the corresponding particle masses, collision coefficients, and hydrodynamic background.

\subsection{Determination of the steady state wall velocity}

The preceding sections provide the ingredients required to determine the steady-state bubble wall velocity. The hydrodynamic analysis determines the temperature and velocity profiles of the plasma for a given trial value of $v_w$, while the microscopic transport calculation determines the perturbations of distribution functions for relevant particles and the associated friction force. We substitute these quantities into the coupled scalar field EOMs and impose two moment conditions to determine the wall velocity and thickness.

For a given wall velocity and wall profile, the EOMs can be written schematically as

\begin{equation}
	\begin{split}
		S_{EOM_\varphi}=&\left(1-v_w^2\right)\frac{d^2\varphi}{dz^2}-\frac{\partial V_{\text{eff}}(\varphi,~\varphi_s,~T_+)}{\partial \varphi} \\
		&-\frac{T_+}{2}\sum_{i}\frac{N_i\partial m_i^2}{\partial\varphi}\left[c_1^i\mu_i+c_2^i(\delta T_i+\delta T_{bg})\right]=0 \,,\\
		S_{EOM_{\varphi_s}}=&\left(1-v_w^2\right)\frac{d^2\varphi_s}{dz^2}-\frac{\partial V_{\text{eff}}(\varphi,~\varphi_s,~T_+)}{\partial \varphi_s} \\
		&-\frac{T_+}{2}\sum_{i}\frac{N_i\partial m_i^2}{\partial\varphi_s}\left[c_1^i\mu_i+c_2^i(\delta T_i+\delta T_{bg})\right]=0 \,,
		\label{eq:final_SEOM}
	\end{split}
\end{equation}
where the temperature entering the effective potential is taken to be the plasma temperature immediately in front of the wall, $T_+$, as determined by the hydrodynamic calculation. The nonequilibrium contribution is expressed in terms of the chemical potential and temperature perturbations obtained from the transport equations.

A direct solution of these coupled equations for both the wall profile and the wall velocity is numerically demanding. Instead, we exploit the steady state conditions satisfied by a wall propagating with constant velocity and constant thickness. We impose two independent moment conditions on the scalar field equations. The first condition corresponds to vanishing net force on the wall,
\begin{equation}
	M_1=\int\,\left[S_{EOM_\varphi}\varphi'+S_{EOM_{\varphi_s}}\varphi_s'\right]\,dz=0 \,,
	\label{eq:stable_condition1}
\end{equation}
and the second condition fixes the wall thickness by requiring the corresponding first-order deformation of the wall profile to vanish,
\begin{equation}
	M_2=\int\,\left[S_{EOM_\varphi}(2\varphi-\varphi_-)\varphi'+S_{EOM_{\varphi_s}}(2\varphi_s-\varphi_{s-}-\varphi_{s+})\varphi_s'\right]\,dz=0 \,.
	\label{eq:stable_condition2}
\end{equation}
These two conditions determine the two unknowns $v_w$ and $L_w$ once the thermodynamic and transport background has been specified.

Several contributions to the moment equations can be evaluated analytically. In particular, the equilibrium part of $M_1$ reduces to the free energy difference between the two phases,
\begin{equation}
	\begin{split}
		&\int\,\left[\left[(1-v_w^2)\frac{d^2\varphi}{dz^2}-\frac{\partial V_{\text{eff}}}{\partial \varphi}\right]\varphi'+\left[(1-v_w^2)\frac{d^2\varphi_s}{dz^2}-\frac{\partial V_{\text{eff}}}{\partial \varphi_s}\right]\varphi_s'\right]\,dz \\
		&=-V_{\text{eff}}(\varphi_-,~\varphi_{s-},~T_+)+V_{\text{eff}}(0,~\varphi_{s+},~T_+)\,,
		\label{eq:M1_constant}
	\end{split}
\end{equation}
while the gradient contribution to the second moment can likewise be evaluated analytically for the adopted $\tanh$ wall profile,
\begin{equation}
	\begin{split}
		&\int\,(1-v_w^2)\frac{d^2\varphi}{dz^2}(2\varphi-\varphi_-)\varphi'+(1-v_w^2)\frac{d^2\varphi_s}{dz^2}(2\varphi_s-\varphi_{s-}-\varphi_{s+})\varphi_s'\,dz \\
		&=\frac{-2(1-v_w^2)[\varphi_-^3+(\varphi_{s-}-\varphi_{s+})^3]}{15 L_w^2} \,.
	\end{split}
\end{equation}
The remaining terms contain the nonequilibrium transport perturbations and are evaluated numerically.

A subtlety arises because the effective potential appearing in the scalar field equations is evaluated at $T_+$, whereas the vacuum expectation values in broken phase entering the phase transition analysis are naturally defined at the temperature $T_-$. In the limit in which $T_--T_+$ is small, the temperature dependence of the  broken phase minimum can be expanded around $T_+$,
\begin{equation}
	\begin{split}
		\frac{\partial V_{\text{eff}}(\varphi_-,~\varphi_{s-},~T_-)}{\partial\varphi_-}&\approx\frac{\partial V_{\text{eff}}(\varphi_-,~\varphi_{s-},~T_+)}{\partial\varphi_-}+\frac{\partial^2 V_{\text{eff}}(\varphi_-,~\varphi_{s-},~T_+)}{\partial\varphi_-\partial T_+}(T_--T_+)=0 \,,\\
		\frac{\partial V_{\text{eff}}(\varphi_-,~\varphi_{s-},~T_-)}{\partial\varphi_{s-}}&\approx\frac{\partial V_{\text{eff}}(\varphi_-,~\varphi_{s-},~T_+)}{\partial\varphi_{s-}}+\frac{\partial^2 V_{\text{eff}}(\varphi_-,~\varphi_{s-},~T_+)}{\partial\varphi_{s-}\partial T_+}(T_--T_+)=0 \,,
		\label{eq:potencial_T_+}
	\end{split}
\end{equation}
and the thermal part of the effective potential satisfies
\begin{equation}
	\begin{split}
		\frac{\partial V_T}{\partial\varphi}&=\sum_i\frac{\partial m_i^2}{\partial \varphi}\int\,\frac{d^3p}{(2\pi)^3}\frac{1}{2E_i}f_{0,i}(x,p) \,,\\
		\frac{\partial V_T}{\partial\varphi_s}&=\sum_i\frac{\partial m_i^2}{\partial \varphi_s}\int\,\frac{d^3p}{(2\pi)^3}\frac{1}{2E_i}f_{0,i}(x,p) \,,
	\end{split}
\end{equation}
which allows the corresponding temperature dependent terms to be related to the coefficients appearing in the transport equations. At large positive $z$, the nonequilibrium perturbations vanish, while the background temperature perturbation remains finite. Consequently, the asymptotic solution of the wall equations does not automatically coincide with the  broken phase minimum evaluated at $T_-$. We therefore determine the asymptotic  broken phase field values self-consistently from the conditions
\begin{equation}
	\begin{split}
		&\frac{\partial V_{\text{eff}}(\varphi_-,~\varphi_{s-},~T_+)}{\partial\varphi_-}+\frac{T_+}{2}\sum_i\frac{\partial m_i^2}{\partial \varphi_-}N_ic_2^i\delta T_{bg}=0 \,,\\
		&\frac{\partial V_{\text{eff}}(\varphi_-,~\varphi_{s-},~T_+)}{\partial\varphi_{s-}}+\frac{T_+}{2}\sum_i\frac{\partial m_i^2}{\partial \varphi_{s-}}N_ic_2^i\delta T_{bg}=0 \,.
	\end{split}
\end{equation}
This procedure ensures that the wall profile satisfies the scalar field equations consistently with the hydrodynamic temperature and the nonequilibrium background.

First, the phase transition parameters are obtained from the finite temperature effective potential. The hydrodynamic equations are then solved to determine the plasma state in front of and behind the wall. The resulting temperature $T_+$ and fluid profiles are used as inputs to the microscopic transport calculation, which provides the nonequilibrium perturbations and hence the frictional contribution to the scalar field equations. Finally, the moment conditions $M_1$ and $M_2$ are evaluated. The procedure is repeated until a point satisfying $M_1=M_2=0$ is found.

\begin{figure}[htbp]
    \centering
    \includegraphics[width=0.98\textwidth]{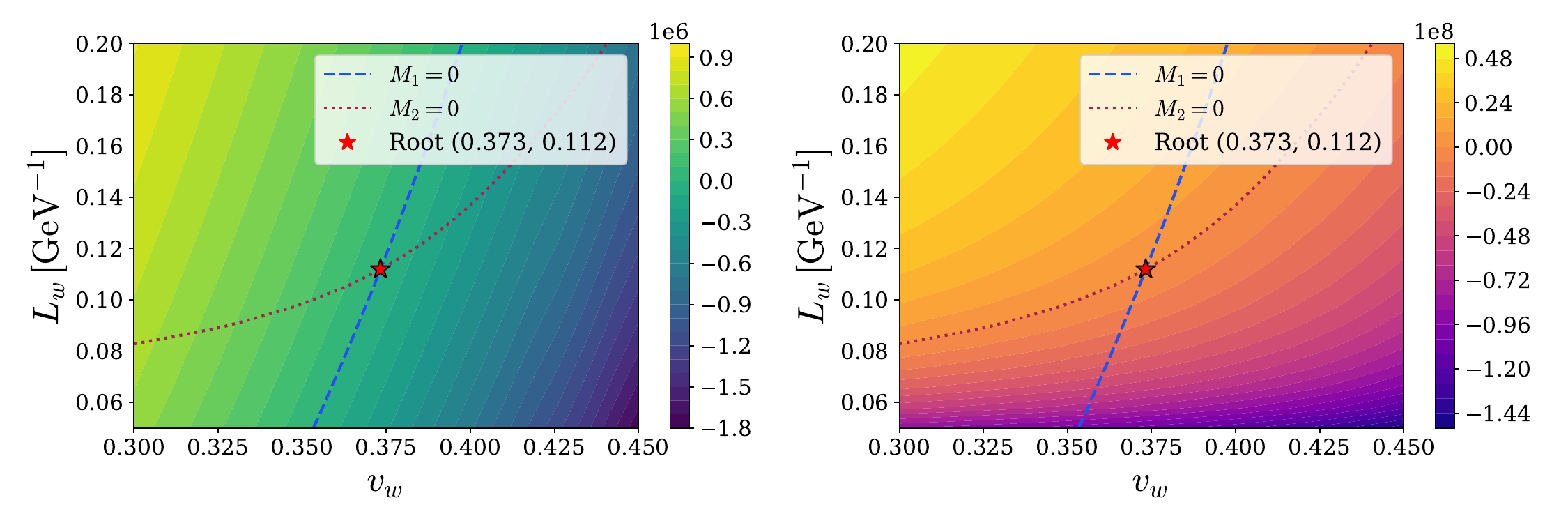}
    \caption{Contours of the moment conditions $M_1$ and $M_2$ in the $(v_w,L_w)$ plane for the benchmark parameter point. 
    The dashed blue and dotted red lines denote $M_1=0$ and $M_2=0$, respectively, with their intersection marked by a red star, corresponding to the steady state solution satisfying $M_1=M_2=0$.}
    \label{fig:deflagration_M_1_M_2_result}
\end{figure}

As a representative example, we first apply the complete procedure to the benchmark parameter point listed in Tab.~\ref{tab:effective_potential_benchmark}. The resulting values of $M_1$ and $M_2$ across the $(v_w,L_w)$ plane are shown in Fig.~\ref{fig:deflagration_M_1_M_2_result}. The intersection of the two zero contours identifies the steady state solution. For the benchmark point considered here, we obtain
\begin{equation}
     v_w=0.373,~~L_w=0.112~[\text{GeV}^{-1}]\,.
\end{equation}

The corresponding value of $L_wT_n$ is approximately $14.5$, indicating a wall that is thick compared with the inverse nucleation temperature and supporting the validity of the semiclassical treatment adopted in the transport calculation. The behavior of the moment conditions also reveals an important limitation of the present approximation. As the trial wall velocity approaches the sound speed, both $M_1$ and $M_2$ develop a rapidly varying behavior. This originates from the hydrodynamic solution: the background temperature and velocity perturbations contain denominators proportional to $3v_w^2-1$, which become singular in the limit $v_w^2\rightarrow c_s^2={1}/{3}$. Therefore, the linearized transport and hydrodynamic treatment employed here becomes unreliable  when the bubble wall velocity approaches the speed of sound. In addition, the analysis in this section is formulated for the deflagration branch, for which the wall velocity is subsonic. Supersonic solutions therefore lie outside the regime of validity of the present calculation.

The benchmark solution above serves as a representative demonstration of the complete numerical procedure. The same procedure is applied to the set of CxSM benchmark points discussed in Sec.~\ref{sec:numerical_results}, where the resulting wall velocities and their dependence on the underlying phase transition strength are analyzed.

\section{Numerical Results}
\label{sec:numerical_results}


Having established the procedure for determining the steady state bubble wall velocity in Sec.~\ref{sec:bubble_wall_dynamics}, we now apply it to a set of benchmark points in the CxSM parameter space. The purpose of this analysis is to investigate how the wall velocity varies with the properties of the underlying FOEWPT. We first specify the parametrization and scanning strategy, and then present the resulting phase transition parameters and bubble wall velocities.

Instead of treating all parameters in the scalar potential as independent inputs, we parametrize the model in terms of the physical quantities $(\varphi,\,\varphi_S,\,m_{h_1},\,m_{h_2},\,m_{\chi},\,\theta,\,a_1)$. The remaining parameters of the scalar potential can then be obtained from the tree-level relations~\cite{Cho:2022zfg} 
\begin{equation}
	\begin{split}
		b_1=-m_{\chi}^2-\frac{\sqrt{2}a_1}{\varphi_S},~&~m^2=-\frac{\lambda}{2}\varphi^2-\frac{\delta_2}{2}\varphi_S^2 \,,\\
		\delta_2=\frac{2}{\varphi\varphi_S}(m_{h_1}^2-m_{h_2}^2)\sin{\theta}\cos{\theta},~&~\lambda=\frac{2}{\varphi^2}(m_{h_1}^2\cos^2{\theta}+m_{h_2}^2\sin^2{\theta}) \,,\\
		b_2=-b_1-\frac{\delta_2}{2}\varphi^2-\frac{d_2}{2}\varphi_S^2-\frac{2\sqrt{2}a_1}{\varphi_S},~&~d_2=\frac{2}{\varphi_S^2}(m_{h_1}^2\sin^2{\theta}+m_{h_2}^2\cos^2{\theta})+\frac{2\sqrt{2}a_1}{\varphi_S^3} \,.
	\end{split}
\end{equation}

This parametrization is convenient because the physical scalar masses, the singlet vacuum expectation value, and the scalar mixing angle can be directly varied while maintaining the desired zero temperature vacuum structure. The benchmark point introduced in Tab.~\ref{tab:effective_potential_benchmark} is therefore more completely specified by the input parameters listed in Tab.~\ref{tab:complet_potential_benchmark}.

\begin{table}[htbp]
\centering
\caption{Input parameters and the corresponding derived parameters for the benchmark point in Tab.~\ref{tab:effective_potential_benchmark}.}
\label{tab:complet_potential_benchmark}
\begin{tabular}{lccccccc}
\hline\hline
Inputs & $\varphi~[\text{GeV}]$ & $\varphi_S~[\text{GeV}]$ & $m_{h_1}~[\text{GeV}]$ & $m_{h_2}~[\text{GeV}]$ & $m_{\chi}~[\text{GeV}]$ & $\theta~[\text{rad}]$ & $a_1$~[GeV$^3$] \\
\hline
  & $246.22$ & $0.6$ & $125.0$ & $124.0$ & $62.5$ & $\pi/3.992$ & $-6576.172$ \\
\hline\hline
Outputs & $m^2$~[GeV$^2$] & $b_1$~[GeV$^2$] & $b_2$~[GeV$^2$] & $\lambda$ & $d_2$ & $\delta_2$ & $a_1$~[GeV$^3$] \\
\hline
  & $-(124.50)^2$ & $(107.67)^2$ & $-(178.00)^2$ & $0.5114$ & $3.9214$ & $1.6855$ & $-6576.172$ \\
\hline
\end{tabular}
\end{table}

For the parameter range considered here, we adopt the nearly degenerate scalar scenario in which $m_{h_1}$ and $m_{h_2}$ are nearly degenerate~\cite{Abe:2021nih, Gross:2017dan}. In this setup, the mixing angle $\theta$ provides a convenient parameter controlling the scalar mixing and, consequently, the structure of the finite temperature effective potential. We therefore vary $\theta$ while keeping the remaining input parameters fixed.

\begin{figure}[htbp]
    \centering
    \includegraphics[width=0.9\textwidth]{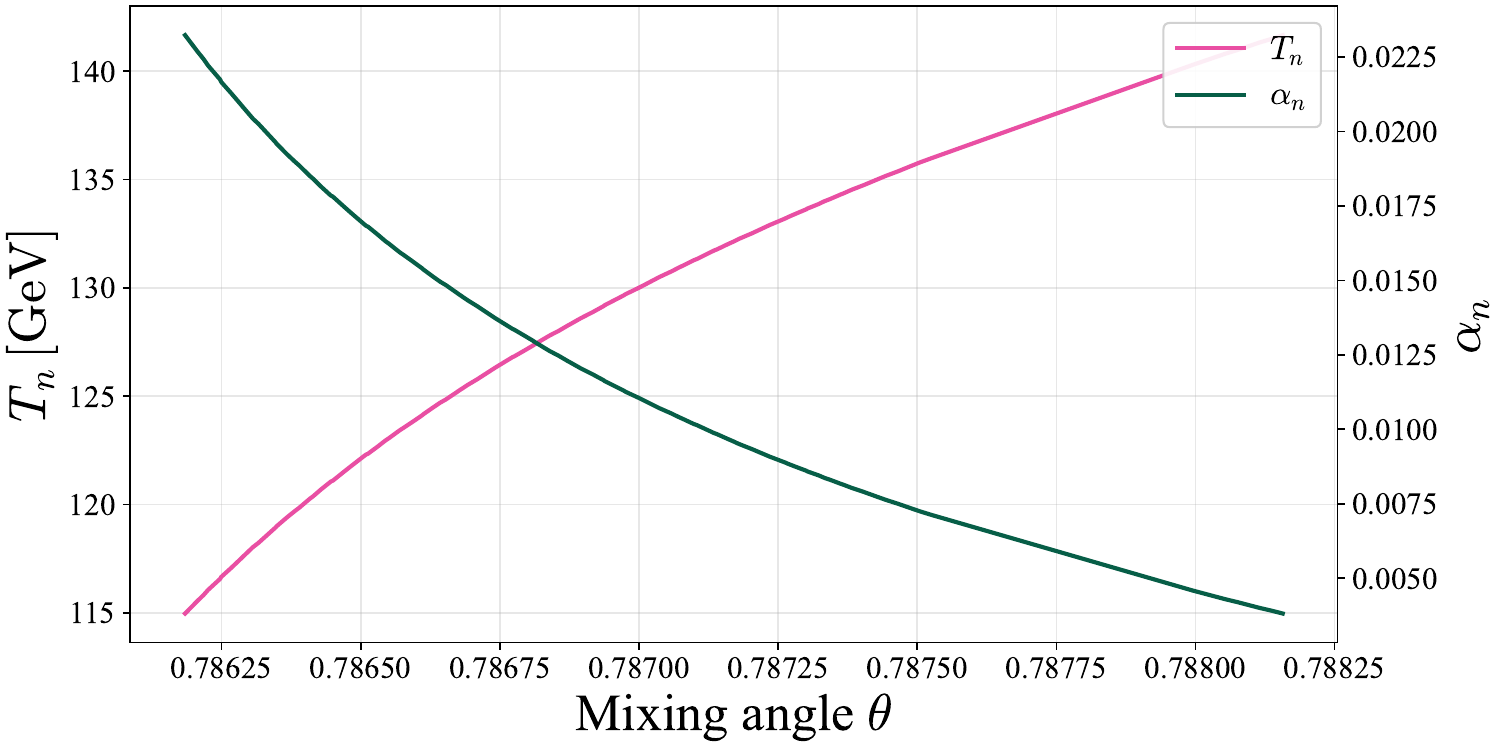}
    \caption{Dependence of the nucleation temperature $T_n$ and transition strength $\alpha_n$ on the mixing angle $\theta$. The pink and green curves show $T_n$ and $\alpha_n$, respectively. Over the parameter range considered, $T_n$ increases while $\alpha_n$ decreases with increasing $\theta$.}
    \label{fig:Tn_alpha_n_with_angle}
\end{figure}

The resulting dependence of the nucleation temperature and phase transition strength on the mixing angle is shown in Fig.~\ref{fig:Tn_alpha_n_with_angle}. Over the range ${\pi}/{4}\leq\theta\leq{\pi}/{3}$, the nucleation temperature $T_n$ increases monotonically with increasing $\theta$, whereas the transition strength $\alpha_n$ decreases with increasing $\theta$. These trends reflect the dependence of the scalar mixing and the finite temperature effective potential on $\theta$.

To restrict the scan to parameter points with a sufficiently 
strong FOEWPT for the phenomenological analysis considered here, we retain only points satisfying $\varphi_c/T_c>1$. This criterion is adopted here as a benchmark selection condition rather than as a model-independent condition for EWBG. The resulting parameter set is then subjected to the full bubble wall calculation developed in Sec.~\ref{sec:bubble_wall_dynamics}. 

\begin{figure}[htbp]
    \centering
    \includegraphics[width=0.9\textwidth]{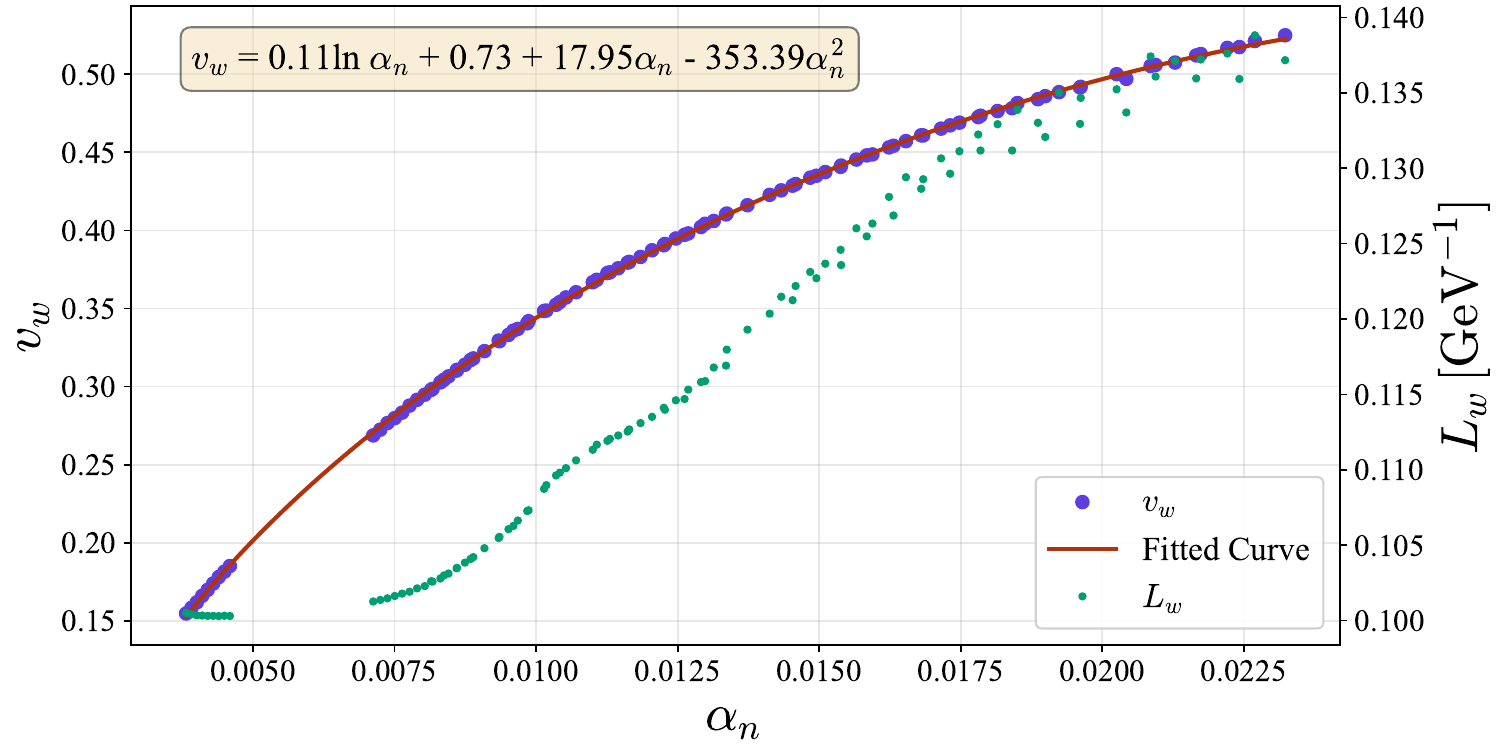}
    \caption{Bubble wall velocity $v_w$ and wall thickness $L_w$ as functions of the transition strength $\alpha_n$. The blue and green points denote the calculated wall velocities and the corresponding wall thickness, respectively, while the red curve shows the empirical fit to the numerical wall velocities. Both $v_w$ and $L_w$ increase with increasing $\alpha_n$, and the fitted curve provides a good description of the calculated wall velocities over the parameter range considered.}
    \label{fig:vw_Lw_with_alphan}
\end{figure}

For each retained parameter point, the hydrodynamic quantities, microscopic collision coefficients, perturbations of the particle distribution function, and moment conditions are recalculated consistently. Solving $M_1(v_w,L_w)=0,\, M_2(v_w,L_w)=0$ then determines the corresponding steady state wall velocity and wall thickness. The resulting wall velocities exhibit a clear dependence on the strength of the phase transition. As shown in Fig.~\ref{fig:vw_Lw_with_alphan}, the bubble wall velocity increases with $\alpha_n$ over the parameter range investigated. This behavior is physically consistent with the increased driving force associated with a stronger phase transition, although the wall velocity is ultimately determined by the competition between the driving force and the microscopic plasma friction discussed in Sec.~\ref{sec:bubble_wall_dynamics}.

Within the scanned parameter region, the numerical results can be described by the empirical fit~\cite{Carena:2025flp} 
\begin{equation}
    v_w=a\ln\alpha_n+b+c\alpha_n+d\alpha_n^2\,,
    \label{eq:Empirical_fitting_function}
\end{equation}
where the coefficients $a$, $b$, $c$, and $d$ are determined from the numerical data. The best fit coefficients are $a\approx0.11,\,b\approx0.73,\,c\approx17.95,$ and $d\approx-353.39$, while the resulting curve is shown together with the numerical results in Fig.~\ref{fig:vw_Lw_with_alphan}. The fit should be regarded as a convenient parametrization of the numerical results within the parameter range considered here, rather than as a universal relation between the wall velocity and the transition strength.

The numerical results demonstrate that the bubble wall velocity is not uniquely fixed by a single conventional phase transition parameter such as $\alpha_n$. Rather, it emerges from the combined effects of the thermodynamic driving force, the hydrodynamic response of the plasma, and the microscopic nonequilibrium friction. The observed correlation between $v_w$ and $\alpha_n$ therefore provides a useful characterization of the present benchmark set, while the underlying calculation retains the full dependence on the CxSM parameters.

\section{Impact of the Calculated Wall Velocity on Baryon Asymmetry}
\label{sec:baryon}

The preceding analysis provides a self-consistent determination of the bubble wall velocity for different CxSM parameter points. Since the wall velocity enters the transport dynamics relevant for EWBG, it is useful to assess how replacing an arbitrarily chosen wall velocity by the value obtained from the bubble wall calculation affects the resulting baryon asymmetry. In this section, we perform a rough estimate using the particle perturbations obtained from the transport calculation.

The baryon to entropy ratio is estimated from
\begin{equation}
	\eta_B=\frac{405\Gamma_{\text{sph}}}{4\pi^2v_wg_*T}\int dz\,\mu_{B_L}f_{\text{sph}}e^{-45\Gamma_{\text{sph}}|z|/(4v_w)}\,,
	\label{eq:eta_B}
\end{equation}
where $\Gamma_{\mathrm{sph}}$ denotes the sphaleron transition rate in the symmetric phase, $f_{\mathrm{sph}}$ parametrizes the suppression of sphaleron processes in the broken phase, and $\mu_{B_L}$ is the effective left-handed baryon chemical potential. As noted in Ref.~\cite{Cline:2020jre}, an additional Lorentz factor $\gamma_w=(1-v_w^2)^{-1/2}$ appears in the denominator of Eq.~\eqref{eq:eta_B}.

We emphasize that the calculation in this section is intended as a phenomenological estimate rather than a complete treatment of EWBG. To isolate the impact of the bubble wall dynamics, we adopt the approximate sphaleron rate $\Gamma_{\mathrm{sph}}\simeq 2.7\times10^{-7}T$ and approximate the sphaleron suppression factor by $1$ in the symmetric phase, while assume that it rapidly vanishes in the broken phase. We estimate the left-handed baryon chemical potential by retaining the quark sector. Rather than solving a separate transport equation for $\mu_{B_L}$, we approximate it in terms of the chemical potential perturbations obtained from our transport calculation as $\mu_{B_L}\simeq0.13(\mu_t-0.5\mu_H)$. A more complete determination of $\mu_{B_L}$ would require treating the left and right handed top quark chemical potentials, $\mu_{t_L}$ and $\mu_{t_R}$, as distinct variables in the transport equations of Eq.~\eqref{eq:transport}.

An important distinction is that the quantities entering this estimate are evaluated for the actual plasma configuration surrounding a steadily propagating wall. In particular, the temperature and fluid velocity immediately in front of the wall are the hydrodynamic quantities $T_+$ and $v_+$ obtained from the calculation in Sec.~\ref{subsec:Hydrodynamics}, rather than the nucleation temperature and an independently chosen fluid velocity. This allows the baryon asymmetry estimate to retain part of the hydrodynamic and microscopic information obtained in the wall velocity calculation.

We first select 36 parameter points uniformly distributed over the accepted mixing angle range discussed in Sec.~\ref{sec:numerical_results}. For each point, the corresponding wall velocity, hydrodynamic plasma state, and particle perturbations obtained in the preceding sections are used to evaluate Eq.~\eqref{eq:eta_B}. This procedure gives a set of baryon asymmetry estimates that consistently incorporates the parameter dependence of the bubble wall dynamics.

For comparison, we construct a conventional phenomenological estimate in which the wall velocity is treated as a freely chosen input. All other quantities entering Eq.~\eqref{eq:eta_B} are first evaluated for the benchmark point in Tab.~\ref{tab:complet_potential_benchmark} and then kept fixed as the wall velocity is varied. The same set of 36 calculated wall velocities is subsequently inserted into Eq.~\eqref{eq:eta_B}. This comparison illustrates the difference between treating the wall velocity as an independent input and using the velocity obtained from the self-consistent wall dynamics calculation.

\begin{figure}[htbp]
    \centering
    \includegraphics[width=0.9\textwidth]{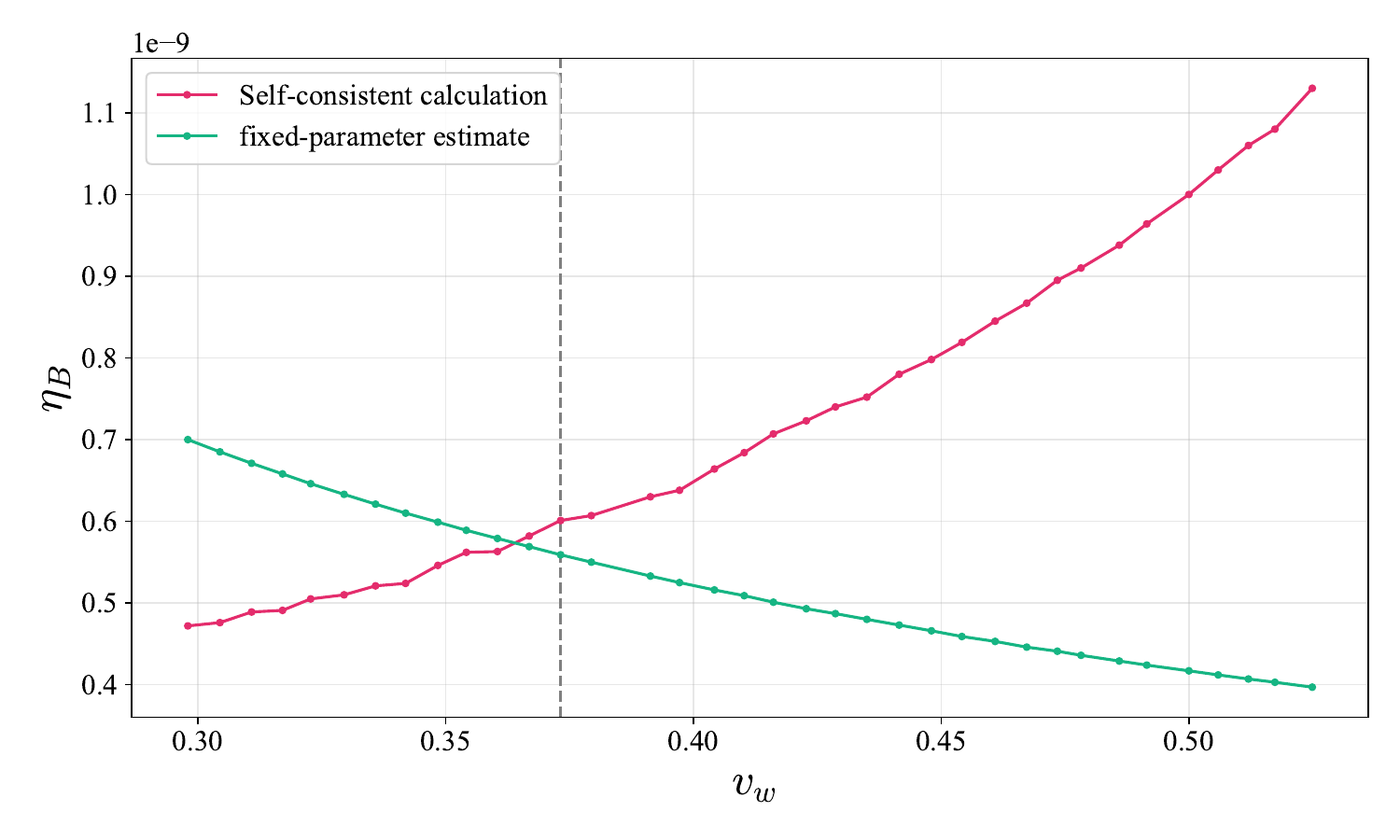}
    \caption{Comparison of the two baryon asymmetry estimates as functions of the bubble wall velocity. The red curve corresponds to the self-consistent estimate using the calculated wall velocity together with the corresponding hydrodynamic and transport quantities. The green curve shows the conventional estimate, in which the wall velocity is varied as an external input while all other quantities in Eq.~\eqref{eq:eta_B} are fixed to those evaluated for the reference benchmark point in Tab.~\ref{tab:complet_potential_benchmark}. The actual wall velocity of the reference parameter set is indicated by a vertical gray line.}
    \label{fig:baryon_asymmetry}
\end{figure}

As shown in Fig.~\ref{fig:baryon_asymmetry}, the two procedures lead to qualitatively different dependences on the wall velocity. The self-consistent estimate increases with the calculated wall velocity over the parameter range considered, whereas the phenomenological estimate obtained by varying only the velocity input decreases. In the latter case, the estimated baryon asymmetry varies approximately within $4\times10^{-10}\lesssim\eta_B\lesssim7\times10^{-10}$, while the value corresponding to the actual wall velocity of the reference parameter set is of order $\eta_B\simeq6\times10^{-10}$. The two curves do not coincide even at the wall velocity corresponding to the reference parameter point. This difference arises because the self-consistent calculation does not simply substitute the wall velocity as an isolated input. Instead, the calculated wall velocity determines the surrounding plasma configuration through the hydrodynamic solution and, in turn, the particle perturbations entering $\mu_{B_L}$. In particular, the fluid velocity immediately in front of the wall, $v_+$, and the temperature $T_+$ are correlated with the wall velocity through the hydrodynamic matching conditions. The conventional estimate neglects these correlations and varies the velocity parameter while keeping the remaining quantities fixed.

The comparison demonstrates that the treatment of the bubble wall velocity can have a non-negligible impact on phenomenological estimates of the baryon asymmetry. In particular, treating $v_w$ as an arbitrary input while neglecting its correlation with the hydrodynamic plasma state and the nonequilibrium particle distributions can lead to a significant shift in the estimated value of $\eta_B$. Although the simplified treatment adopted here does not constitute a complete calculation of EWBG, it illustrates the phenomenological importance of determining the wall velocity from the underlying phase transition dynamics rather than fixing it independently.

\section{Conclusions and discussions}
\label{sec:conclusion}

We have calculated the steady-state bubble wall velocity for a specific phase transition pattern in the CxSM in which both the Higgs and singlet backgrounds vary across the wall, while the singlet background remains nonzero in both phases. Because some particle masses vary with the coupled background fields, the microscopic friction must be included in both scalar field EOMs. We combine the Boltzmann transport treatment of the heavy particle species with the hydrodynamic response of the plasma, and determine the wall velocity and thickness self-consistently from the two moment conditions, $M_1=M_2=0$. This extends the transport plus hydrodynamics framework to this specific coupled scalar fields phase transition pattern.

For the benchmark point in Tab.~\ref{tab:effective_potential_benchmark}, we obtain $v_w=0.373$ and $L_w=0.112~\mathrm{GeV}^{-1}$. The value $L_wT_n\simeq14.5$ supports the applicability of the thick wall assumption adopted in our semiclassical transport treatment. In the nearly degenerate scalar scenario, both $v_w$ and $L_w$ increase with the phase transition strength $\alpha_n$ over the parameter range studied. An empirical fit describes the calculated velocities within this scan, but it should not be interpreted as a universal relation between $v_w$ and $\alpha_n$. The wall velocity is determined by the interplay between the thermodynamic driving force, microscopic friction, and the hydrodynamic response of the plasma.

Nevertheless, our results remain subject to several uncertainties. A more robust determination of the bubble wall velocity will require improvements in at least three aspects, which we outline below. We leave a detailed investigation of these issues to future work.

We further estimate the baryon asymmetry by using the calculated wall velocities and the nonequilibrium chemical potential perturbations obtained from the transport calculation. Over the parameter range considered, this self-consistent estimate differs from the result obtained by varying $v_w$ alone while keeping the other inputs fixed at a reference benchmark point. This difference reflects the correlations between $v_w$, the fluid velocity $v_+$, the temperature $T_+$, and the particle perturbations. The comparison illustrates why the wall velocity should be determined together with the plasma response rather than treated as an independent input. However, our estimate of the baryon asymmetry is intended as a phenomenological illustration rather than a complete prediction of electroweak baryogenesis. A more comprehensive calculation would require a consistent treatment of CP-violating sources and charge transport.

The wall velocity depends on the finite temperature effective potential. We employ a one-loop potential with daisy resummation and a fixed renormalization scale. Higher-order corrections, as well as the gauge and renormalization scale dependence of the potential can affect $T_n$, $\alpha_n$, and hence $v_w$. To assess these effects, one could vary the renormalization scale and compare our results with RG-improved thermodynamics based on dimensional reduction to a three-dimensional effective field theory (3D EFT). A consistent power counting is especially important for radiatively induced phase transitions. These issues have been studied in the CxSM and in analyses of perturbative uncertainties in other models~\cite{Lofgren:2021ogg,Hirvonen:2021zej,Schicho:2022wty,Biekotter:2025npc,Lewicki:2024xan,Kierkla:2026bnm}. The impact of loop corrections on the bubble wall dynamics should also be studied further~\cite{Dashko:2024anp}.

At the macroscopic level, we use an ideal fluid description, a bag equation of state, and linearized perturbations of the background temperature and fluid velocity. The resulting equations become singular as $v_w$ approaches the sound speed, $c_s=1/\sqrt{3}$. The origin of this behavior has been debated, and several alternative treatments have been proposed~\cite{laurent_fluid_2020,dorsch_sonic_2022,Laurent:2022jrs,Dorsch:2024jjl}. These include modified fluid equations that remain regular near and above the sound speed, as well as treatments that retain the spatial dependence of the background fluid. At the microscopic level, our friction calculation is based on a linearized Boltzmann equation. Higher moment truncations, spectral methods for solving the Boltzmann equation, and treatments of nonlinear terms in the distribution functions have also been developed~\cite{Dorsch:2023tss,DeCurtis:2023hil,DeCurtis:2024hvh}. The assumed $\tanh$ wall profiles are another approximation. Recent studies have identified the wall profile as a source of uncertainty, although its effect on $v_w$ remains to be quantified~\cite{Laurent:2022jrs,Friedlander:2020tnq,vandeVis:2025plm}. Our analysis is restricted to subsonic deflagrations. The LTE and ballistic limits provide complementary bounds on $v_w$, while WallGo offers a way to cross check the numerical results~\cite{Ai:2024btx,ekstedt_how_2024}.

A steady-state solution does not establish whether the bubble wall can reach this state during its evolution. Hydrodynamic simulations and recent studies of the time dependent pressure on the wall suggest that the heating wave may form after the wall has already started to accelerate~\cite{Krajewski:2024gma,Laurent:2026jvx}. The steady-state approximation may therefore be inadequate in some cases. The nonequilibrium quantum field theory provides an alternative approach to describing the interaction between the wall and the plasma~\cite{Ai:2025bjw}. Solving the time dependent equations for the coupled scalar fields and plasma would help determine whether the steady-state solutions obtained here can be reached dynamically.

\begin{acknowledgments}
We thank Philipp Schicho for insightful comments and 
Siyu Jiang for helpful discussions. This work is supported by the National Natural Science Foundation of China (NNSFC)
Grant No.12475111 and the Fundamental Research Funds for the Central Universities, Sun Yat-sen University.
\end{acknowledgments}

\appendix

\section{Scattering matrix elements}
\label{app:scattering_matrix}

In this appendix, we collect the dominant $2\rightarrow2$ scattering processes and the corresponding squared matrix elements used in the evaluation of the collision terms in Sec.~\ref{subsubsec:collision_terms}. Only the channels retained in the numerical calculation are listed.

For the top quark, we include the following dominant processes:
\begin{itemize}
	\item $t+g\rightarrow t+g$,~~~$|\mathcal{M}|^2=-\frac{128}{3}g_s^4\frac{su}{(u-m_{t,T}^2)^2}+96g_s^4\frac{s^2+u^2}{(t-m_{t,T}^2)^2}$
	\item $t+q(\bar{q})\rightarrow t+q(\bar{q})$,~~~$|\mathcal{M}|^2=160g_s^4\frac{s^2+u^2}{(t-m_{t,T}^2)^2}$
	\item $t+\bar{t}\rightarrow g+g$,~~~$|\mathcal{M}|^2=\frac{128}{3}g_s^4\left(\frac{u}{t-m_{t,T}^2}+\frac{t}{u-m_{t,T}^2}\right)$
	\item $t+\bar{t}\rightarrow H+g$,~~~$|\mathcal{M}|^2=8y_t^2g_s^2\left(\frac{u}{t-m_{t,T}^2}+\frac{t}{u-m_{t,T}^2}\right)$
	\item $t+g\rightarrow t+H$,~~~$|\mathcal{M}|^2=-8y_t^2g_s^2\frac{s}{t-m_{t,T}^2}$
\end{itemize}
For the gauge boson sector, the dominant processes retained in the calculation are
\begin{itemize}
	\item $W+f\rightarrow W+f$,~~~$|\mathcal{M}|^2=360g_2^4\frac{u^2}{(t-m_{W,T}^2)^2}-\frac{27}{2}g_2^4\left(\frac{3s}{u-m_{q,T}^2}+\frac{s}{t-m_{l,T}^2}\right)$
	\item $W+q\rightarrow q+g$,~~~$|\mathcal{M}|^2=-72g_s^2g_2^2\frac{s}{t-m_{q,T}^2}$
	\item $W+g\rightarrow q+\bar{q}$,~~~$|\mathcal{M}|^2=-72g_s^2g_2^2\frac{s}{t-m_{q,T}^2}$
	\item $W+W\rightarrow f+\bar{f}$,~~~$|\mathcal{M}|^2=-\frac{27}{2}g_2^4\left(\frac{3s}{t-m_{q,T}^2}+\frac{t}{u-m_{l,T}^2}\right)$
\end{itemize}

For the scalar sector, the dominant scattering channels involving $H$ and $\chi$ are related by inverse reactions. We therefore list each independent matrix element only once:
\begin{itemize}
	\item $H+H\rightarrow\chi+\chi$,~~~$|\mathcal{M}|^2=\frac{(\delta_2^2\varphi^2+d_2^2\varphi_S^2)^2}{16}\left[\frac{1}{(t-m_{\chi,T}^2)}+\frac{1}{(u-m_{\chi,T}^2)}\right]^2$
	
	\item $H+\chi\rightarrow \chi+H$,
	\[
	\begin{split}
		|\mathcal{M}|^2 = &\frac{(\delta_2^2\varphi^2+d_2^2\varphi_S^2)^2}{16(u-m_{\chi,T}^2)^2}+\frac{\varphi^4}{16(t-m_{H,T}^2)^2} \\
		&\times\left[(3\lambda_{111}D_1+\lambda_{112}D_2)^2+2(\lambda_{112}D_1+\lambda_{122}D_2)^2+(\lambda_{122}D_1+3\lambda_{222}D_2)^2\right]
	\end{split}
	\]
\end{itemize}
where the same squared matrix element is used for both directions of the reaction.

\section{Phase space integration for collision terms}
\label{app:phase_space_integration}

The momentum moments of the collision terms introduced in Sec.~\ref{subsubsec:collision_terms} involve multidimensional phase space integrals. In their original form, these integrals are computationally expensive. We therefore reduce the dimensionality of the phase space integrals before performing the numerical calculation.

For a generic $2\rightarrow2$ scattering process, the phase space integration can be written as
\begin{equation}
	\int dR=\int\frac{d^3pd^3kd^3p'd^3k'}{(2\pi)^{12}16E_pE_kE_{p'}E_{k'}}(2\pi)^4\delta^{(4)}(p+k-p'-k')\,.
\end{equation}
Under the ultrarelativistic approximation $E_p,E_k,E_{p'},E_{k'}\rightarrow p,k,p',k'$, the 12-dimensional integral can be reduced to a five-dimensional integral,
\begin{equation}
	\int dR=\frac{1}{8(2\pi)^6}\int_0^{+\infty}\,dq\int_{-q}^{+q}\,d\omega\int_{\frac{q-\omega}{2}}^{+\infty}\,dp\int_{\frac{q+\omega}{2}}^{+\infty}\,dk\int_0^{2\pi}\,d\phi \,,
\end{equation}
where the Mandelstam variables can be expressed in terms of the integration variables as
\begin{equation}
	\begin{split}
		p'&=p+\omega,~~~k'=k-\omega,~~~t=\omega^2-q^2,~~~u=-t-s \,,\\
		s&=\frac{-t}{2q^2}\biggl\{[(p+p')(k+k')+q^2]-\cos\phi\sqrt{(4pp'+t)(4kk'+t)}\biggr\} \,.
	\end{split}
\end{equation}

A similar reduction can be performed for annihilation processes. The corresponding phase space integral is reduced to five dimensions,
\begin{equation}
	\int dR=\frac{1}{8(2\pi)^6}\int_0^{+\infty}\,d\omega\int_{0}^{\omega}\,dq\int_{\frac{\omega+q}{2}}^{\frac{\omega-q}{2}}\,dp\int_{\frac{\omega+q}{2}}^{\frac{\omega-q}{2}}\,dp'\int_0^{2\pi}\,d\phi \,,
\end{equation}
with the Mandelstam variables given by
\begin{equation}
	\begin{split}
		k&=\omega-p,~~~k'=\omega-p',~~~s=\omega^2-q^2,~~~u=-s-t \,,\\
		t&=\frac{s}{2q^2}\biggl\{[(p-k)(p'-k')-q^2]+\cos\phi\sqrt{(4pk-s)(4p'k'-s)}\biggr\} \,.
	\end{split}
\end{equation}
After this reduction, the collision coefficients entering the transport equations take the generic form
\begin{equation}
	\Gamma_{\delta,i}=\frac{1}{2N_i}\int dR\,|\mathcal{M}|^2f_{0i}f_{0j}(1\mp f_{0m})(1\mp f_{0n})\times\mathcal{W}_{\delta} \,,
\end{equation}
where $\mathcal{W}_{\delta}$ denotes the weight function associated with the corresponding momentum moment. For scattering processes, the relevant weight functions are
\begin{equation}
	\begin{split}
		\Gamma_{\mu1,i}\rightarrow\mathcal{W}_{\mu1}=0\times\frac{1}{T^3}\,,~~~&\Gamma_{T1,i}\rightarrow\mathcal{W}_{T1}=\frac{p-p'}{T^4}\,,\\
		\Gamma_{\mu2,i}\rightarrow\mathcal{W}_{\mu2}=0\times\frac{p}{T^4}\,,~~~&\Gamma_{T2,i}\rightarrow\mathcal{W}_{T2}=\frac{p(p-p')}{T^5}\,,\\
		\Gamma_{v,i}\rightarrow\mathcal{W}_{v}&=\frac{p(p-p')}{3T^5} \,,
	\end{split}
\end{equation}
while for annihilation processes they are
\begin{equation}
	\begin{split}
		\Gamma_{\mu1,i}\rightarrow\mathcal{W}_{\mu1}=\frac{2}{T^3}\,,~~~&\Gamma_{T1,i}\rightarrow\mathcal{W}_{T1}=\frac{p+k}{T^4} \,,\\
		\Gamma_{\mu2,i}\rightarrow\mathcal{W}_{\mu2}=\frac{p}{T^4}\,,~~~&\Gamma_{T2,i}\rightarrow\mathcal{W}_{T2}=\frac{p(p+k)}{T^5} \,,\\
		\Gamma_{v,i}\rightarrow\mathcal{W}_{v}&=\frac{p(p+k)}{3T^5} \,.
	\end{split}
\end{equation}
The remaining five-dimensional integrals are evaluated numerically using Monte Carlo integration. This dimensional reduction substantially improves the computational efficiency of the collision term calculation and makes the procedure practical for the parameter scans discussed in the main text.

\section{Numerical collision coefficients}
\label{app:numerical_collsion}

Using the phase space reduction and numerical integration procedure described above, we evaluate the collision coefficients entering the transport equations. For the benchmark setup considered in the main text, the resulting coefficients are summarized in Tab.~\ref{tab:collision_coefficients}. They are presented in units of the plasma temperature $T$, such that the numerical prefactors are dimensionless.

The coefficients are obtained by summing over the dominant scattering channels specified in App.~\ref{app:scattering_matrix}. The effective particle species considered in the transport calculation are the top quark $t$, the effective gauge boson species $W$, the degenerate scalar species $H$, and the singlet scalar $\chi$.

\begin{table}[htbp]
\centering
\caption{Numerical collision coefficients for the benchmark setup. The coefficients are given in units of $T$.}
\label{tab:collision_coefficients}
\begin{tabular}{lccccc}
\hline\hline
Species
& $\Gamma_{\mu1}/T$
& $\Gamma_{T1}/T$
& $\Gamma_{\mu2}/T$
& $\Gamma_{T2}/T$
& $\Gamma_v/T$ \\
\hline
$t$
& $9.01\times10^{-4}$
& $2.09\times10^{-3}$
& $2.07\times10^{-3}$
& $8.93\times10^{-3}$
& $2.98\times10^{-3}$ \\

$W$
& $1.05\times10^{-3}$
& $2.23\times10^{-3}$
& $2.12\times10^{-3}$
& $5.88\times10^{-3}$
& $1.96\times10^{-3}$ \\

$H$
& $2.84\times10^{-1}$
& $9.23\times10^{-3}$
& $9.24\times10^{-3}$
& $8.98\times10^{-3}$
& $2.99\times10^{-3}$ \\

$\chi$
& $5.70\times10^{-1}$
& $1.85\times10^{-2}$
& $1.85\times10^{-2}$
& $1.79\times10^{-2}$
& $5.97\times10^{-3}$ \\
\hline\hline
\end{tabular}
\end{table}

\bibliography{v_w_reference}

@article{Moore:1995si,
    author = "Moore, Guy D. and Prokopec, Tomislav",
    title = "{How fast can the wall move? A Study of the electroweak phase transition dynamics}",
    eprint = "hep-ph/9506475",
    archivePrefix = "arXiv",
    reportNumber = "PUPT-1544, PUP-TH-1544, LANCS-TH-9517",
    doi = "10.1103/PhysRevD.52.7182",
    journal = "Phys. Rev. D",
    volume = "52",
    pages = "7182--7204",
    year = "1995"
}

@article{vandeVis:2025efm,
    author = "van de Vis, Jorinde and de Vries, Jordy and Postma, Marieke",
    title = "{Bubble trouble: A review on electroweak baryogenesis}",
    eprint = "2508.09989",
    archivePrefix = "arXiv",
    primaryClass = "hep-ph",
    reportNumber = "CERN-TH-2025-161, Nikhef 2025-012",
    doi = "10.1016/j.ppnp.2026.104244",
    journal = "Prog. Part. Nucl. Phys.",
    volume = "150",
    pages = "104244",
    year = "2026"
}

@article{espinosa_energy_2010,
    author = "Espinosa, Jose R. and Konstandin, Thomas and No, Jose M. and Servant, Geraldine",
    title = "{Energy Budget of Cosmological First-order Phase Transitions}",
    eprint = "1004.4187",
    archivePrefix = "arXiv",
    primaryClass = "hep-ph",
    reportNumber = "CERN-PH-TH-2010-027",
    doi = "10.1088/1475-7516/2010/06/028",
    journal = "JCAP",
    volume = "06",
    pages = "028",
    year = "2010"
}

@article{Konstandin:2014zta,
    author = "Konstandin, Thomas and Nardini, Germano and Rues, Ingo",
    title = "{From Boltzmann equations to steady wall velocities}",
    eprint = "1407.3132",
    archivePrefix = "arXiv",
    primaryClass = "hep-ph",
    reportNumber = "DESY-14-127, NSF-KITP-14-089",
    doi = "10.1088/1475-7516/2014/09/028",
    journal = "JCAP",
    volume = "09",
    pages = "028",
    year = "2014"
}

@article{Morrissey:2012db,
    author = "Morrissey, David E. and Ramsey-Musolf, Michael J.",
    title = "{Electroweak baryogenesis}",
    eprint = "1206.2942",
    archivePrefix = "arXiv",
    primaryClass = "hep-ph",
    reportNumber = "NPAC-12-08",
    doi = "10.1088/1367-2630/14/12/125003",
    journal = "New J. Phys.",
    volume = "14",
    pages = "125003",
    year = "2012"
}

@article{cho_electroweak_2021,
    author = "Cho, Gi-Chol and Idegawa, Chikako and Senaha, Eibun",
    title = "{Electroweak phase transition in a complex singlet extension of the Standard Model with degenerate scalars}",
    eprint = "2105.11830",
    archivePrefix = "arXiv",
    primaryClass = "hep-ph",
    reportNumber = "OCHA-PP-366",
    doi = "10.1016/j.physletb.2021.136787",
    journal = "Phys. Lett. B",
    volume = "823",
    pages = "136787",
    year = "2021"
}

@article{grinstein_electroweak_2008,
    author = "Grinstein, Benjamin and Trott, Michael",
    title = "{Electroweak Baryogenesis with a Pseudo-Goldstone Higgs}",
    eprint = "0806.1971",
    archivePrefix = "arXiv",
    primaryClass = "hep-ph",
    reportNumber = "UCSD-PTH-08-04",
    doi = "10.1103/PhysRevD.78.075022",
    journal = "Phys. Rev. D",
    volume = "78",
    pages = "075022",
    year = "2008"
}

@article{John:2000zq,
    author = "John, P. and Schmidt, M. G.",
    title = "{Do stops slow down electroweak bubble walls?}",
    eprint = "hep-ph/0002050",
    archivePrefix = "arXiv",
    reportNumber = "HE-THEP-00-04",
    doi = "10.1016/S0550-3213(00)00768-9",
    journal = "Nucl. Phys. B",
    volume = "598",
    pages = "291--305",
    year = "2001",
    note = "[Erratum: Nucl.Phys.B 648, 449--452 (2003)]"
}

@article{laurent_fluid_2020,
    author = "Laurent, Benoit and Cline, James M.",
    title = "{Fluid equations for fast-moving electroweak bubble walls}",
    eprint = "2007.10935",
    archivePrefix = "arXiv",
    primaryClass = "hep-ph",
    doi = "10.1103/PhysRevD.102.063516",
    journal = "Phys. Rev. D",
    volume = "102",
    number = "6",
    pages = "063516",
    year = "2020"
}

@article{ekstedt_how_2024,
    author = "Ekstedt, Andreas and Gould, Oliver and Hirvonen, Joonas and Laurent, Benoit and Niemi, Lauri and Schicho, Philipp and van de Vis, Jorinde",
    title = "{How fast does the WallGo? A package for computing wall velocities in first-order phase transitions}",
    eprint = "2411.04970",
    archivePrefix = "arXiv",
    primaryClass = "hep-ph",
    reportNumber = "CERN-TH-2024-174, DESY-24-162, HIP-2024-21/TH",
    doi = "10.1007/JHEP04(2025)101",
    journal = "JHEP",
    volume = "04",
    pages = "101",
    year = "2025"
}

@article{konstandin_hydrodynamic_2011,
    author = "Konstandin, Thomas and No, Jose M.",
    title = "{Hydrodynamic obstruction to bubble expansion}",
    eprint = "1011.3735",
    archivePrefix = "arXiv",
    primaryClass = "hep-ph",
    doi = "10.1088/1475-7516/2011/02/008",
    journal = "JCAP",
    volume = "02",
    pages = "008",
    year = "2011"
}

@article{wang_hydrodynamic_2023,
    author = "Wang, Shao-Jiang and Yuwen, Zi-Yan",
    title = "{Hydrodynamic backreaction force of cosmological bubble expansion}",
    eprint = "2205.02492",
    archivePrefix = "arXiv",
    primaryClass = "hep-ph",
    doi = "10.1103/PhysRevD.107.023501",
    journal = "Phys. Rev. D",
    volume = "107",
    number = "2",
    pages = "023501",
    year = "2023"
}

@article{cline_baryogenesis_2021,
    author = "Cline, James M. and Friedlander, Avi and He, Dong-Ming and Kainulainen, Kimmo and Laurent, Benoit and Tucker-Smith, David",
    title = "{Baryogenesis and gravity waves from a UV-completed electroweak phase transition}",
    eprint = "2102.12490",
    archivePrefix = "arXiv",
    primaryClass = "hep-ph",
    doi = "10.1103/PhysRevD.103.123529",
    journal = "Phys. Rev. D",
    volume = "103",
    number = "12",
    pages = "123529",
    year = "2021"
}

@article{kozaczuk_bubble_2015,
    author = "Kozaczuk, Jonathan",
    title = "{Bubble Expansion and the Viability of Singlet-Driven Electroweak Baryogenesis}",
    eprint = "1506.04741",
    archivePrefix = "arXiv",
    primaryClass = "hep-ph",
    doi = "10.1007/JHEP10(2015)135",
    journal = "JHEP",
    volume = "10",
    pages = "135",
    year = "2015"
}

@article{steinhardt_relativistic_1982,
    author = "Steinhardt, Paul Joseph",
    title = "{Relativistic Detonation Waves and Bubble Growth in False Vacuum Decay}",
    reportNumber = "UPR-0181T",
    doi = "10.1103/PhysRevD.25.2074",
    journal = "Phys. Rev. D",
    volume = "25",
    pages = "2074",
    year = "1982"
}

@article{dorsch_wall_2021,
    author = "Dorsch, Glauber C. and Huber, Stephan J. and Konstandin, Thomas",
    title = "{On the wall velocity dependence of electroweak baryogenesis}",
    eprint = "2106.06547",
    archivePrefix = "arXiv",
    primaryClass = "hep-ph",
    reportNumber = "DESY-21-089, DESY 21-089",
    doi = "10.1088/1475-7516/2021/08/020",
    journal = "JCAP",
    volume = "08",
    pages = "020",
    year = "2021"
}

@article{dorsch_sonic_2022,
    author = "Dorsch, Glauber C. and Huber, Stephan J. and Konstandin, Thomas",
    title = "{A sonic boom in bubble wall friction}",
    eprint = "2112.12548",
    archivePrefix = "arXiv",
    primaryClass = "hep-ph",
    reportNumber = "DESY 21-224",
    doi = "10.1088/1475-7516/2022/04/010",
    journal = "JCAP",
    volume = "04",
    number = "04",
    pages = "010",
    year = "2022"
}

@article{Dorsch:2023tss,
    author = "Dorsch, Glauber C. and Pinto, Daniel A.",
    title = "{Bubble wall velocities with an extended fluid Ansatz}",
    eprint = "2312.02354",
    archivePrefix = "arXiv",
    primaryClass = "hep-ph",
    doi = "10.1088/1475-7516/2024/04/027",
    journal = "JCAP",
    volume = "04",
    pages = "027",
    year = "2024"
}

@article{Laurent:2022jrs,
    author = "Laurent, Benoit and Cline, James M.",
    title = "{First principles determination of bubble wall velocity}",
    eprint = "2204.13120",
    archivePrefix = "arXiv",
    primaryClass = "hep-ph",
    doi = "10.1103/PhysRevD.106.023501",
    journal = "Phys. Rev. D",
    volume = "106",
    number = "2",
    pages = "023501",
    year = "2022"
}

@article{Megevand:2009gh,
    author = "Megevand, Ariel and Sanchez, Alejandro D.",
    title = "{Velocity of electroweak bubble walls}",
    eprint = "0908.3663",
    archivePrefix = "arXiv",
    primaryClass = "hep-ph",
    doi = "10.1016/j.nuclphysb.2009.09.019",
    journal = "Nucl. Phys. B",
    volume = "825",
    pages = "151--176",
    year = "2010"
}

@article{Krajewski:2024gma,
    author = "Krajewski, Tomasz and Lewicki, Marek and Zych, Mateusz",
    title = "{Bubble-wall velocity in local thermal equilibrium: hydrodynamical simulations vs analytical treatment}",
    eprint = "2402.15408",
    archivePrefix = "arXiv",
    primaryClass = "astro-ph.CO",
    doi = "10.1007/JHEP05(2024)011",
    journal = "JHEP",
    volume = "05",
    pages = "011",
    year = "2024"
}

@article{Ghosh:2025rbt,
    author = "Ghosh, Dilip Kumar and Mukherjee, Debadrita and Mukherjee, Koustav and Pramanick, Rohan",
    title = "{Complex scalar singlet model: Electroweak phase transition and gravitational waves}",
    eprint = "2511.13426",
    archivePrefix = "arXiv",
    primaryClass = "hep-ph",
    doi = "10.1103/q89d-gggz",
    journal = "Phys. Rev. D",
    volume = "114",
    number = "5",
    pages = "055021",
    year = "2026"
}

@article{Yang:2024npd,
    author = "Yang, Aidi and Huang, Fa Peng",
    title = "{Detectability of the phase transition gravitational waves in the DFSZ axion model}",
    eprint = "2404.18703",
    archivePrefix = "arXiv",
    primaryClass = "hep-ph",
    doi = "10.1088/1475-7516/2025/05/028",
    journal = "JCAP",
    volume = "05",
    pages = "028",
    year = "2025"
}

@article{Huang:2023eal,
    author = "Huang, Fa Peng",
    title = "{Electroweak phase transition and gravitational wave: Higgs potential from 10{\ensuremath{-}}11 s of early universe to 2022}",
    doi = "10.1016/j.nuclphysbps.2023.04.013",
    journal = "Nucl. Part. Phys. Proc.",
    volume = "330-335",
    pages = "21--26",
    year = "2023"
}

@article{Chang:2025rda,
    author = "Chang, Yijie and Tang, Shihang and Deng, Haowen and Wang, Yefeng and Ding, Ran and Huang, Fa Peng",
    title = "{Effects of Early-Universe Inhomogeneity on Bubble Formation: Primordial Black Holes as an Extreme Case}",
    eprint = "2511.11408",
    archivePrefix = "arXiv",
    primaryClass = "hep-ph",
    month = "11",
    year = "2025"
}

@article{Yang:2025ybx,
    author = "Yang, Aidi and Idegawa, Chikako and Huang, Fa Peng",
    title = "{Model parameter reconstruction of electroweak phase transition with TianQin and LISA: Insights from the dimension-six model}",
    eprint = "2511.02612",
    archivePrefix = "arXiv",
    primaryClass = "hep-ph",
    doi = "10.1103/dbz8-hg6r",
    journal = "Phys. Rev. D",
    volume = "113",
    number = "11",
    pages = "115053",
    year = "2026"
}

@article{Qiu:2025tmn,
    author = "Qiu, Dayun and Jiang, Siyu and Huang, Fa Peng",
    title = "{A new source of phase transition gravitational waves: heavy particle braking across bubble walls}",
    eprint = "2508.04314",
    archivePrefix = "arXiv",
    primaryClass = "hep-ph",
    doi = "10.1007/JHEP02(2026)112",
    journal = "JHEP",
    volume = "02",
    pages = "112",
    year = "2026"
}

@article{Barger:2008jx,
    author = "Barger, Vernon and Langacker, Paul and McCaskey, Mathew and Ramsey-Musolf, Michael and Shaughnessy, Gabe",
    title = "{Complex Singlet Extension of the Standard Model}",
    eprint = "0811.0393",
    archivePrefix = "arXiv",
    primaryClass = "hep-ph",
    reportNumber = "MADPH-08-1516, NUHEP-TH-08-06, ANL-HEP-PR-08-58, NPAC-08-21",
    doi = "10.1103/PhysRevD.79.015018",
    journal = "Phys. Rev. D",
    volume = "79",
    pages = "015018",
    year = "2009"
}

@article{Idegawa:2023bkh,
    author = "Idegawa, Chikako and Senaha, Eibun",
    title = "{Electron electric dipole moment and electroweak baryogenesis in a complex singlet extension of the Standard Model with degenerate scalars}",
    eprint = "2309.09430",
    archivePrefix = "arXiv",
    primaryClass = "hep-ph",
    reportNumber = "OCHA-PP-377",
    doi = "10.1016/j.physletb.2023.138332",
    journal = "Phys. Lett. B",
    volume = "848",
    pages = "138332",
    year = "2024"
}

@article{Coleman:1973jx,
    author = "Coleman, Sidney R. and Weinberg, Erick J.",
    title = "{Radiative Corrections as the Origin of Spontaneous Symmetry Breaking}",
    doi = "10.1103/PhysRevD.7.1888",
    journal = "Phys. Rev. D",
    volume = "7",
    pages = "1888--1910",
    year = "1973"
}

@article{Wainwright:2011kj,
    author = "Wainwright, Carroll L.",
    title = "{CosmoTransitions: Computing Cosmological Phase Transition Temperatures and Bubble Profiles with Multiple Fields}",
    eprint = "1109.4189",
    archivePrefix = "arXiv",
    primaryClass = "hep-ph",
    doi = "10.1016/j.cpc.2012.04.004",
    journal = "Comput. Phys. Commun.",
    volume = "183",
    pages = "2006--2013",
    year = "2012"
}

@article{Anderson:1991zb,
    author = "Anderson, Greg W. and Hall, Lawrence J.",
    title = "{The Electroweak phase transition and baryogenesis}",
    reportNumber = "LBL-31169, UCB-PTH-91-41",
    doi = "10.1103/PhysRevD.45.2685",
    journal = "Phys. Rev. D",
    volume = "45",
    pages = "2685--2698",
    year = "1992"
}

@article{Chiang:2018gsn,
    author = "Chiang, Cheng-Wei and Li, Yen-Ting and Senaha, Eibun",
    title = "{Revisiting electroweak phase transition in the standard model with a real singlet scalar}",
    eprint = "1808.01098",
    archivePrefix = "arXiv",
    primaryClass = "hep-ph",
    reportNumber = "CTPU-PTC-18-24",
    doi = "10.1016/j.physletb.2018.12.017",
    journal = "Phys. Lett. B",
    volume = "789",
    pages = "154--159",
    year = "2019"
}

@article{Luo:2025ewp,
    author = "Luo, Jun and others",
    title = "{Fundamental physics and cosmology with TianQin}",
    eprint = "2502.20138",
    archivePrefix = "arXiv",
    primaryClass = "gr-qc",
    doi = "10.1007/s41114-025-00063-2",
    journal = "Living Rev. Rel.",
    volume = "29",
    number = "1",
    pages = "1",
    year = "2026"
}

@article{TianQin:2015yph,
    author = "Luo, Jun and others",
    collaboration = "TianQin",
    title = "{TianQin: a space-borne gravitational wave detector}",
    eprint = "1512.02076",
    archivePrefix = "arXiv",
    primaryClass = "astro-ph.IM",
    doi = "10.1088/0264-9381/33/3/035010",
    journal = "Class. Quant. Grav.",
    volume = "33",
    number = "3",
    pages = "035010",
    year = "2016"
}

@article{LISA:2017pwj,
    author = "Amaro-Seoane, Pau and others",
    collaboration = "LISA",
    title = "{Laser Interferometer Space Antenna}",
    eprint = "1702.00786",
    archivePrefix = "arXiv",
    primaryClass = "astro-ph.IM",
    month = "2",
    year = "2017"
}

@article{Hu:2017mde,
    author = "Hu, Wen-Rui and Wu, Yue-Liang",
    title = "{The Taiji Program in Space for gravitational wave physics and the nature of gravity}",
    doi = "10.1093/nsr/nwx116",
    journal = "Natl. Sci. Rev.",
    volume = "4",
    number = "5",
    pages = "685--686",
    year = "2017"
}

@article{BarrosoMancha:2020fay,
    author = "Barroso Mancha, Marc and Prokopec, Tomislav and Swiezewska, Bogumila",
    title = "{Field-theoretic derivation of bubble-wall force}",
    eprint = "2005.10875",
    archivePrefix = "arXiv",
    primaryClass = "hep-th",
    doi = "10.1007/JHEP01(2021)070",
    journal = "JHEP",
    volume = "01",
    pages = "070",
    year = "2021"
}

@article{Gross:2017dan,
    author = "Gross, Christian and Lebedev, Oleg and Toma, Takashi",
    title = "{Cancellation Mechanism for Dark-Matter{\textendash}Nucleon Interaction}",
    eprint = "1708.02253",
    archivePrefix = "arXiv",
    primaryClass = "hep-ph",
    reportNumber = "HIP-2017-20-TH, TUM-HEP-1091-17, HIP-2017-20/TH, TUM-HEP/1091/17",
    doi = "10.1103/PhysRevLett.119.191801",
    journal = "Phys. Rev. Lett.",
    volume = "119",
    number = "19",
    pages = "191801",
    year = "2017"
}

@article{Cho:2022zfg,
    author = "Cho, Gi-Chol and Idegawa, Chikako and Sugihara, Rio",
    title = "{A complex singlet extension of the standard model and multi-critical point principle}",
    eprint = "2212.13029",
    archivePrefix = "arXiv",
    primaryClass = "hep-ph",
    reportNumber = "OCHA-PP-374",
    doi = "10.1016/j.physletb.2023.137757",
    journal = "Phys. Lett. B",
    volume = "839",
    pages = "137757",
    year = "2023"
}

@article{Abe:2021nih,
    author = "Abe, Sachiho and Cho, Gi-Chol and Mawatari, Kentarou",
    title = "{Probing a degenerate-scalar scenario in a pseudoscalar dark-matter model}",
    eprint = "2101.04887",
    archivePrefix = "arXiv",
    primaryClass = "hep-ph",
    reportNumber = "OCHA-PP-362",
    doi = "10.1103/PhysRevD.104.035023",
    journal = "Phys. Rev. D",
    volume = "104",
    number = "3",
    pages = "035023",
    year = "2021"
}

@article{Kuzmin:1985mm,
    author = "Kuzmin, V. A. and Rubakov, V. A. and Shaposhnikov, M. E.",
    title = "{On the Anomalous Electroweak Baryon Number Nonconservation in the Early Universe}",
    reportNumber = "IC/85/8",
    doi = "10.1016/0370-2693(85)91028-7",
    journal = "Phys. Lett. B",
    volume = "155",
    pages = "36",
    year = "1985"
}

@article{Funakubo:1996dw,
    author = "Funakubo, Koichi",
    title = "{CP violation and baryogenesis at the electroweak phase transition}",
    eprint = "hep-ph/9608358",
    archivePrefix = "arXiv",
    reportNumber = "SAGA-HE-108",
    doi = "10.1143/PTP.96.475",
    journal = "Prog. Theor. Phys.",
    volume = "96",
    pages = "475--520",
    year = "1996"
}

@article{Kajantie:1996mn,
    author = "Kajantie, K. and Laine, M. and Rummukainen, K. and Shaposhnikov, Mikhail E.",
    title = "{Is there a~ hot electroweak phase transition at $m_H \gtrsim m_W$?}",
    eprint = "hep-ph/9605288",
    archivePrefix = "arXiv",
    reportNumber = "CERN-TH-96-126, HD-THEP-96-15, IUHET-333",
    doi = "10.1103/PhysRevLett.77.2887",
    journal = "Phys. Rev. Lett.",
    volume = "77",
    pages = "2887--2890",
    year = "1996"
}

@article{Csikor:1998eu,
    author = "Csikor, F. and Fodor, Z. and Heitger, J.",
    title = "{Endpoint of the hot electroweak phase transition}",
    eprint = "hep-ph/9809291",
    archivePrefix = "arXiv",
    reportNumber = "ITP-BUDAPEST-541, KEK-TH-580, KEK-PREPRINT-98-160, MS-TPI-98-16",
    doi = "10.1103/PhysRevLett.82.21",
    journal = "Phys. Rev. Lett.",
    volume = "82",
    pages = "21--24",
    year = "1999"
}

@article{Rubakov:1996vz,
    author = "Rubakov, V. A. and Shaposhnikov, M. E.",
    title = "{Electroweak baryon number nonconservation in the early universe and in high-energy collisions}",
    eprint = "hep-ph/9603208",
    archivePrefix = "arXiv",
    reportNumber = "CERN-TH-96-13, INR-0913-96",
    doi = "10.1070/PU1996v039n05ABEH000145",
    journal = "Usp. Fiz. Nauk",
    volume = "166",
    pages = "493--537",
    year = "1996"
}

@article{Dine:1992wr,
    author = "Dine, Michael and Leigh, Robert G. and Huet, Patrick Y. and Linde, Andrei D. and Linde, Dmitri A.",
    title = "{Towards the theory of the electroweak phase transition}",
    eprint = "hep-ph/9203203",
    archivePrefix = "arXiv",
    reportNumber = "SLAC-PUB-5741, SCIPP-92-07, SU-ITP-92-7",
    doi = "10.1103/PhysRevD.46.550",
    journal = "Phys. Rev. D",
    volume = "46",
    pages = "550--571",
    year = "1992"
}

@article{Liu:1992tn,
    author = "Liu, Bao-Hua and McLerran, Larry D. and Turok, Neil",
    title = "{Bubble nucleation and growth at a baryon number producing electroweak phase transition}",
    reportNumber = "TPI-MINN-92-18-T",
    doi = "10.1103/PhysRevD.46.2668",
    journal = "Phys. Rev. D",
    volume = "46",
    pages = "2668--2688",
    year = "1992"
}

@article{Ignatius:1993qn,
    author = "Ignatius, J. and Kajantie, K. and Kurki-Suonio, H. and Laine, M.",
    title = "{The growth of bubbles in cosmological phase transitions}",
    eprint = "astro-ph/9309059",
    archivePrefix = "arXiv",
    reportNumber = "HU-TFT-93-43",
    doi = "10.1103/PhysRevD.49.3854",
    journal = "Phys. Rev. D",
    volume = "49",
    pages = "3854--3868",
    year = "1994"
}

@article{Moore:1995ua,
    author = "Moore, Guy D. and Prokopec, Tomislav",
    title = "{Bubble wall velocity in a first order electroweak phase transition}",
    eprint = "hep-ph/9503296",
    archivePrefix = "arXiv",
    reportNumber = "PUPT-1531, LANCASTER-TH-9503, PUP-TH-1531-(1995), LANCASTER-TH-9503-(1995)",
    doi = "10.1103/PhysRevLett.75.777",
    journal = "Phys. Rev. Lett.",
    volume = "75",
    pages = "777--780",
    year = "1995"
}

@article{Cline:2020jre,
    author = "Cline, James M. and Kainulainen, Kimmo",
    title = "{Electroweak baryogenesis at high bubble wall velocities}",
    eprint = "2001.00568",
    archivePrefix = "arXiv",
    primaryClass = "hep-ph",
    reportNumber = "CERN-TH-2019-227",
    doi = "10.1103/PhysRevD.101.063525",
    journal = "Phys. Rev. D",
    volume = "101",
    number = "6",
    pages = "063525",
    year = "2020"
}

@article{Carena:2025flp,
    author = "Carena, Marcela and Ireland, Aurora and Ou, Tong and Wang, Isaac R.",
    title = "{The discriminant power of bubble wall velocities: gravitational waves and electroweak baryogenesis}",
    eprint = "2504.17841",
    archivePrefix = "arXiv",
    primaryClass = "hep-ph",
    reportNumber = "FERMILAB-PUB-25-0266-T",
    doi = "10.1007/JHEP09(2025)175",
    journal = "JHEP",
    volume = "09",
    pages = "175",
    year = "2025"
}

@article{Wang:2020zlf,
    author = "Wang, Xiao and Huang, Fa Peng and Zhang, Xinmin",
    title = "{Bubble wall velocity beyond leading-log approximation in electroweak phase transition}",
    eprint = "2011.12903",
    archivePrefix = "arXiv",
    primaryClass = "hep-ph",
    month = "11",
    year = "2020"
}

@article{Kosowsky:1991ua,
    author = "Kosowsky, Arthur and Turner, Michael S. and Watkins, Richard",
    title = "{Gravitational Radiation from Colliding Vacuum Bubbles}",
    reportNumber = "FERMILAB-PUB-91-323-A",
    doi = "10.1103/PhysRevD.45.4514",
    journal = "Phys. Rev. D",
    volume = "45",
    pages = "4514--4535",
    year = "1992"
}

@article{Kosowsky:1992vn,
    author = "Kosowsky, Arthur and Turner, Michael S.",
    title = "{Gravitational radiation from colliding vacuum bubbles: envelope approximation to many bubble collisions}",
    eprint = "astro-ph/9211004",
    archivePrefix = "arXiv",
    reportNumber = "FERMILAB-PUB-92-295-A",
    doi = "10.1103/PhysRevD.47.4372",
    journal = "Phys. Rev. D",
    volume = "47",
    pages = "4372--4391",
    year = "1993"
}

@article{Kamionkowski:1993fg,
    author = "Kamionkowski, Marc and Kosowsky, Arthur and Turner, Michael S.",
    title = "{Gravitational radiation from first order phase transitions}",
    eprint = "astro-ph/9310044",
    archivePrefix = "arXiv",
    reportNumber = "IASSNS-HEP-93-44, FERMILAB-PUB-93-235-A",
    doi = "10.1103/PhysRevD.49.2837",
    journal = "Phys. Rev. D",
    volume = "49",
    pages = "2837--2851",
    year = "1994"
}

@article{Hindmarsh:2013xza,
    author = "Hindmarsh, Mark and Huber, Stephan J. and Rummukainen, Kari and Weir, David J.",
    title = "{Gravitational waves from the sound of a first order phase transition}",
    eprint = "1304.2433",
    archivePrefix = "arXiv",
    primaryClass = "hep-ph",
    reportNumber = "HIP-2013-07-TH",
    doi = "10.1103/PhysRevLett.112.041301",
    journal = "Phys. Rev. Lett.",
    volume = "112",
    pages = "041301",
    year = "2014"
}

@article{Baker:2019ndr,
    author = "Baker, Michael J. and Kopp, Joachim and Long, Andrew J.",
    title = "{Filtered Dark Matter at a First Order Phase Transition}",
    eprint = "1912.02830",
    archivePrefix = "arXiv",
    primaryClass = "hep-ph",
    doi = "10.1103/PhysRevLett.125.151102",
    journal = "Phys. Rev. Lett.",
    volume = "125",
    number = "15",
    pages = "151102",
    year = "2020"
}

@article{Chway:2019kft,
    author = "Chway, Dongjin and Jung, Tae Hyun and Shin, Chang Sub",
    title = "{Dark matter filtering-out effect during a first-order phase transition}",
    eprint = "1912.04238",
    archivePrefix = "arXiv",
    primaryClass = "hep-ph",
    reportNumber = "CTPU-19-37",
    doi = "10.1103/PhysRevD.101.095019",
    journal = "Phys. Rev. D",
    volume = "101",
    number = "9",
    pages = "095019",
    year = "2020"
}

@article{Huang:2017kzu,
    author = "Huang, Fa Peng and Li, Chong Sheng",
    title = "{Probing the baryogenesis and dark matter relaxed in phase transition by gravitational waves and colliders}",
    eprint = "1709.09691",
    archivePrefix = "arXiv",
    primaryClass = "hep-ph",
    reportNumber = "CTPU-17-34",
    doi = "10.1103/PhysRevD.96.095028",
    journal = "Phys. Rev. D",
    volume = "96",
    number = "9",
    pages = "095028",
    year = "2017"
}

@article{Ai:2025bjw,
    author = "Ai, Wen-Yuan and Carosi, Matthias and Garbrecht, Bjorn and Tamarit, Carlos and Vanvlasselaer, Miguel",
    title = "{Bubble wall dynamics from nonequilibrium quantum field theory}",
    eprint = "2504.13725",
    archivePrefix = "arXiv",
    primaryClass = "hep-ph",
    reportNumber = "MITP-25-029",
    doi = "10.1007/JHEP08(2025)077",
    journal = "JHEP",
    volume = "08",
    pages = "077",
    year = "2025"
}

@article{Arnold:1993wc,
    author = "Arnold, Peter Brockway",
    title = "{One loop fluctuation - dissipation formula for bubble wall velocity}",
    eprint = "hep-ph/9302258",
    archivePrefix = "arXiv",
    reportNumber = "UW-PT-93-2",
    doi = "10.1103/PhysRevD.48.1539",
    journal = "Phys. Rev. D",
    volume = "48",
    pages = "1539--1545",
    year = "1993"
}

@article{Caprini:2019egz,
    author = "Caprini, Chiara and others",
    title = "{Detecting gravitational waves from cosmological phase transitions with LISA: an update}",
    eprint = "1910.13125",
    archivePrefix = "arXiv",
    primaryClass = "astro-ph.CO",
    reportNumber = "DESY-19-159, IPPP/19/27, HIP-2019-14/TH, MITP/19-066, IFT-UAM/CSIC-19-139",
    doi = "10.1088/1475-7516/2020/03/024",
    journal = "JCAP",
    volume = "03",
    pages = "024",
    year = "2020"
}

@article{Moore:2000wx,
    author = "Moore, Guy D.",
    title = "{Electroweak bubble wall friction: Analytic results}",
    eprint = "hep-ph/0001274",
    archivePrefix = "arXiv",
    reportNumber = "UW-PT-00-04",
    doi = "10.1088/1126-6708/2000/03/006",
    journal = "JHEP",
    volume = "03",
    pages = "006",
    year = "2000"
}

@article{Dorsch:2018pat,
    author = "Dorsch, Glauber C. and Huber, Stephan J. and Konstandin, Thomas",
    title = "{Bubble wall velocities in the Standard Model and beyond}",
    eprint = "1809.04907",
    archivePrefix = "arXiv",
    primaryClass = "hep-ph",
    reportNumber = "DESY 18-162, DESY-18-162",
    doi = "10.1088/1475-7516/2018/12/034",
    journal = "JCAP",
    volume = "12",
    pages = "034",
    year = "2018"
}

@article{Friedlander:2020tnq,
    author = "Friedlander, Avi and Banta, Ian and Cline, James M. and Tucker-Smith, David",
    title = "{Wall speed and shape in singlet-assisted strong electroweak phase transitions}",
    eprint = "2009.14295",
    archivePrefix = "arXiv",
    primaryClass = "hep-ph",
    doi = "10.1103/PhysRevD.103.055020",
    journal = "Phys. Rev. D",
    volume = "103",
    number = "5",
    pages = "055020",
    year = "2021"
}

@article{vandeVis:2025plm,
    author = "van de Vis, Jorinde and Schicho, Philipp and Niemi, Lauri and Laurent, Benoit and Hirvonen, Joonas and Gould, Oliver",
    title = "{WallGo investigates: Theoretical uncertainties in the bubble wall velocity}",
    eprint = "2510.27691",
    archivePrefix = "arXiv",
    primaryClass = "hep-ph",
    reportNumber = "CERN-TH-2025-221",
    doi = "10.1007/JHEP04(2026)041",
    journal = "JHEP",
    volume = "04",
    pages = "041",
    year = "2026"
}

@article{Hindmarsh:2020hop,
    author = {Hindmarsh, Mark B. and L{\"u}ben, Marvin and Lumma, Johannes and Pauly, Martin},
    title = "{Phase transitions in the early universe}",
    eprint = "2008.09136",
    archivePrefix = "arXiv",
    primaryClass = "astro-ph.CO",
    reportNumber = "MPP-2020-163, HIP-2020-27/TH",
    doi = "10.21468/SciPostPhysLectNotes.24",
    journal = "SciPost Phys. Lect. Notes",
    volume = "24",
    pages = "1",
    year = "2021"
}

@article{Athron:2023xlk,
    author = "Athron, Peter and Bal{\'a}zs, Csaba and Fowlie, Andrew and Morris, Lachlan and Wu, Lei",
    title = "{Cosmological phase transitions: From perturbative particle physics to gravitational waves}",
    eprint = "2305.02357",
    archivePrefix = "arXiv",
    primaryClass = "hep-ph",
    doi = "10.1016/j.ppnp.2023.104094",
    journal = "Prog. Part. Nucl. Phys.",
    volume = "135",
    pages = "104094",
    year = "2024"
}

@article{Megevand:2013hwa,
    author = "M{\'e}gevand, Ariel",
    title = "{Friction forces on phase transition fronts}",
    eprint = "1303.4233",
    archivePrefix = "arXiv",
    primaryClass = "astro-ph.CO",
    doi = "10.1088/1475-7516/2013/07/045",
    journal = "JCAP",
    volume = "07",
    pages = "045",
    year = "2013"
}

@article{Baker:2021nyl,
    author = "Baker, Michael J. and Breitbach, Moritz and Kopp, Joachim and Mittnacht, Lukas",
    title = "{Primordial black holes from first-order cosmological phase transitions}",
    eprint = "2105.07481",
    archivePrefix = "arXiv",
    primaryClass = "astro-ph.CO",
    reportNumber = "CERN-TH-2021-079, MITP-21-023",
    doi = "10.1016/j.physletb.2025.139625",
    journal = "Phys. Lett. B",
    volume = "868",
    pages = "139625",
    year = "2025"
}

@article{Baker:2021sno,
    author = "Baker, Michael J. and Breitbach, Moritz and Kopp, Joachim and Mittnacht, Lukas",
    title = "{Detailed calculation of primordial black hole formation during first-order cosmological phase transitions}",
    eprint = "2110.00005",
    archivePrefix = "arXiv",
    primaryClass = "astro-ph.CO",
    reportNumber = "CERN-TH-2021-113, MITP-21-036",
    doi = "10.1103/PhysRevD.111.063544",
    journal = "Phys. Rev. D",
    volume = "111",
    number = "6",
    pages = "063544",
    year = "2025"
}

@article{Ai:2024btx,
    author = "Ai, Wen-Yuan and Laurent, Benoit and van de Vis, Jorinde",
    title = "{Bounds on the bubble wall velocity}",
    eprint = "2411.13641",
    archivePrefix = "arXiv",
    primaryClass = "hep-ph",
    reportNumber = "CERN-TH-2024-198, KCL-PH-TH/2024-57",
    doi = "10.1007/JHEP02(2025)119",
    journal = "JHEP",
    volume = "02",
    pages = "119",
    year = "2025"
}

@article{Bodeker:2009qy,
    author = "Bodeker, Dietrich and Moore, Guy D.",
    title = "{Can electroweak bubble walls run away?}",
    eprint = "0903.4099",
    archivePrefix = "arXiv",
    primaryClass = "hep-ph",
    doi = "10.1088/1475-7516/2009/05/009",
    journal = "JCAP",
    volume = "05",
    pages = "009",
    year = "2009"
}

@article{Biekotter:2025npc,
    author = {Biek{\"o}tter, Thomas and Dashko, Andrii and L{\"o}schner, Maximilian and Weiglein, Georg},
    title = "{Perturbative aspects of the electroweak phase transition with a complex singlet and implications for gravitational wave predictions}",
    eprint = "2511.14831",
    archivePrefix = "arXiv",
    primaryClass = "hep-ph",
    reportNumber = "DESY-25-131, IFT-UAM/CSIC-25-104",
    doi = "10.1007/JHEP10(2026)060",
    journal = "JHEP",
    volume = "10",
    pages = "060",
    year = "2026"
}

@article{Dashko:2024anp,
    author = "Dashko, Andrii and Ekstedt, Andreas",
    title = "{Bubble-wall speed with loop corrections}",
    eprint = "2411.05075",
    archivePrefix = "arXiv",
    primaryClass = "hep-ph",
    reportNumber = "DESY-24-169",
    doi = "10.1007/JHEP03(2025)024",
    journal = "JHEP",
    volume = "03",
    pages = "024",
    year = "2025"
}

@article{Branchina:2026nql,
    author = "Branchina, Carlo and De Curtis, Stefania and Delle Rose, Luigi and Notari, Alessio and Panico, Giuliano and Starbuck, Matthew",
    title = "{Thermal Masses and Bubble-Wall Friction in Cosmological Phase Transitions}",
    eprint = "2607.14867",
    archivePrefix = "arXiv",
    primaryClass = "hep-ph",
    reportNumber = "KA-TP-15-2026, P3H-26-048",
    month = "7",
    year = "2026"
}

@article{Branchina:2025adj,
    author = "Branchina, Carlo and Conaci, Angela and Delle Rose, Luigi and De Curtis, Stefania",
    title = "{Bubble wall velocity with out-of-equilibrium corrections}",
    eprint = "2510.21942",
    archivePrefix = "arXiv",
    primaryClass = "hep-ph",
    doi = "10.1103/nmkw-7kgk",
    journal = "Phys. Rev. D",
    volume = "113",
    number = "3",
    pages = "035024",
    year = "2026"
}

@article{Branchina:2025jou,
    author = "Branchina, Carlo and Conaci, Angela and Delle Rose, Luigi and De Curtis, Stefania",
    title = "{Electroweak phase transition and bubble wall velocity in local thermal equilibrium}",
    eprint = "2504.21213",
    archivePrefix = "arXiv",
    primaryClass = "hep-ph",
    doi = "10.1103/lkht-1nv1",
    journal = "Phys. Rev. D",
    volume = "112",
    number = "9",
    pages = "095008",
    year = "2025"
}

@article{DeCurtis:2024hvh,
    author = "De Curtis, Stefania and Delle Rose, Luigi and Guiggiani, Andrea and Gil Muyor, {\'A}ngel and Panico, Giuliano",
    title = "{Non-linearities in cosmological bubble wall dynamics}",
    eprint = "2401.13522",
    archivePrefix = "arXiv",
    primaryClass = "hep-ph",
    doi = "10.1007/JHEP05(2024)009",
    journal = "JHEP",
    volume = "05",
    pages = "009",
    year = "2024"
}

@article{DeCurtis:2023hil,
    author = "De Curtis, Stefania and Delle Rose, Luigi and Guiggiani, Andrea and Gil Muyor, {\'A}ngel and Panico, Giuliano",
    title = "{Collision integrals for cosmological phase transitions}",
    eprint = "2303.05846",
    archivePrefix = "arXiv",
    primaryClass = "hep-ph",
    doi = "10.1007/JHEP05(2023)194",
    journal = "JHEP",
    volume = "05",
    pages = "194",
    year = "2023"
}

@article{DeCurtis:2022hlx,
    author = "De Curtis, Stefania and Rose, Luigi Delle and Guiggiani, Andrea and Muyor, {\'A}ngel Gil and Panico, Giuliano",
    title = "{Bubble wall dynamics at the electroweak phase transition}",
    eprint = "2201.08220",
    archivePrefix = "arXiv",
    primaryClass = "hep-ph",
    doi = "10.1007/JHEP03(2022)163",
    journal = "JHEP",
    volume = "03",
    pages = "163",
    year = "2022"
}

@article{Azatov:2021ifm,
    author = "Azatov, Aleksandr and Vanvlasselaer, Miguel and Yin, Wen",
    title = "{Dark Matter production from relativistic bubble walls}",
    eprint = "2101.05721",
    archivePrefix = "arXiv",
    primaryClass = "hep-ph",
    reportNumber = "SISSA 03/2021/FISI",
    doi = "10.1007/JHEP03(2021)288",
    journal = "JHEP",
    volume = "03",
    pages = "288",
    year = "2021"
}

@article{Azatov:2022tii,
    author = "Azatov, Aleksandr and Barni, Giulio and Chakraborty, Sabyasachi and Vanvlasselaer, Miguel and Yin, Wen",
    title = "{Ultra-relativistic bubbles from the simplest Higgs portal and their cosmological consequences}",
    eprint = "2207.02230",
    archivePrefix = "arXiv",
    primaryClass = "hep-ph",
    reportNumber = "SISSA 12/2022/FISI TU-1157",
    doi = "10.1007/JHEP10(2022)017",
    journal = "JHEP",
    volume = "10",
    pages = "017",
    year = "2022"
}

@article{Lofgren:2021ogg,
    author = {L{\"o}fgren, Johan and Ramsey-Musolf, Michael J. and Schicho, Philipp and Tenkanen, Tuomas V. I.},
    title = "{Nucleation at Finite Temperature: A Gauge-Invariant Perturbative Framework}",
    eprint = "2112.05472",
    archivePrefix = "arXiv",
    primaryClass = "hep-ph",
    reportNumber = "ACFI-T21-15, HIP-2021-44/TH, NORDITA 2021-110",
    doi = "10.1103/PhysRevLett.130.251801",
    journal = "Phys. Rev. Lett.",
    volume = "130",
    number = "25",
    pages = "251801",
    year = "2023"
}

@article{Hirvonen:2021zej,
    author = {Hirvonen, Joonas and L{\"o}fgren, Johan and Ramsey-Musolf, Michael J. and Schicho, Philipp and Tenkanen, Tuomas V. I.},
    title = "{Computing the gauge-invariant bubble nucleation rate in finite temperature effective field theory}",
    eprint = "2112.08912",
    archivePrefix = "arXiv",
    primaryClass = "hep-ph",
    reportNumber = "ACFI-T21-16, HIP-2021-45/TH, NORDITA 2021-111",
    doi = "10.1007/JHEP07(2022)135",
    journal = "JHEP",
    volume = "07",
    pages = "135",
    year = "2022"
}

@article{Schicho:2022wty,
    author = "Schicho, Philipp and Tenkanen, Tuomas V. I. and White, Graham",
    title = "{Combining thermal resummation and gauge invariance for electroweak phase transition}",
    eprint = "2203.04284",
    archivePrefix = "arXiv",
    primaryClass = "hep-ph",
    reportNumber = "HIP-2022-2/TH, NORDITA 2022-009",
    doi = "10.1007/JHEP11(2022)047",
    journal = "JHEP",
    volume = "11",
    pages = "047",
    year = "2022"
}

@article{Lewicki:2024xan,
    author = "Lewicki, Marek and Merchand, Marco and Sagunski, Laura and Schicho, Philipp and Schmitt, Daniel",
    title = "{Impact of theoretical uncertainties on model parameter reconstruction from GW signals sourced by cosmological phase transitions}",
    eprint = "2403.03769",
    archivePrefix = "arXiv",
    primaryClass = "hep-ph",
    doi = "10.1103/PhysRevD.110.023538",
    journal = "Phys. Rev. D",
    volume = "110",
    number = "2",
    pages = "023538",
    year = "2024"
}

@article{Kierkla:2026bnm,
    author = "Kierkla, Maciej and Lewicki, Marek and Schicho, Philipp and Schmitt, Daniel and Swiezewska, Bogumila",
    title = "{Theoretical uncertainties in reconstructing model parameters with gravitational waves from supercooled phase transitions}",
    eprint = "2607.18233",
    archivePrefix = "arXiv",
    primaryClass = "hep-ph",
    reportNumber = "KA-TP-17-2026",
    month = "7",
    year = "2026"
}

@article{Dorsch:2024jjl,
    author = "Dorsch, Gl{\'a}uber C. and Konstandin, Thomas and Perboni, Enrico and Pinto, Daniel A.",
    title = "{Non-singular solutions to the Boltzmann equation with a fluid Ansatz}",
    eprint = "2412.09266",
    archivePrefix = "arXiv",
    primaryClass = "hep-ph",
    reportNumber = "DESY-24-193",
    doi = "10.1088/1475-7516/2025/04/033",
    journal = "JCAP",
    volume = "04",
    pages = "033",
    year = "2025"
}

@article{Laurent:2026jvx,
    author = "Laurent, Benoit and Vanvlasselaer, Miguel",
    title = "{Dynamical evolution of the pressure on the bubble wall}",
    eprint = "2606.30740",
    archivePrefix = "arXiv",
    primaryClass = "hep-ph",
    month = "6",
    year = "2026"
}

\end{document}